\documentclass[12pt,oneside, a4paper]{article}

\pdfoutput=1  
 
\ifx\pdfoutput\undefined
\usepackage[dvips,bookmarks=false]{hyperref}	
\else
\usepackage{hyperref}	
\fi
\hypersetup{colorlinks,bookmarksopen,bookmarksnumbered,citecolor=blue,linkcolor=blue,pdfstartview=FitH,urlcolor=blue}

\usepackage{graphicx}
\usepackage{subfigure}
\usepackage{amssymb}
\usepackage{cite}
\usepackage{bm}
\usepackage{amsmath,amsthm}
\usepackage{cleveref}
\usepackage{siunitx}
\usepackage{slashed}
\usepackage{multirow}
\usepackage{multicol}
\numberwithin{equation}{section}

\usepackage[normalem]{ulem}

\newcommand{\capdef}{}
\newcommand{\mycaption}[2][\capdef]{\renewcommand{\capdef}{#2}%
       \caption[#1]{{\footnotesize #2}}}

\newcommand{\delCP}{\delta_{\rm CP}}

\begin{document}

\begin{titlepage}

\begin{center}

\vspace*{2cm}
        {\Large\bf T versus CP effects in DUNE and T2HK} 
\vspace{1cm}

\renewcommand{\thefootnote}{\fnsymbol{footnote}}
{\bf Sabya Sachi Chatterjee}$^a$\footnote[1]{sabya.chatterjee@kit.edu},
{\bf Sudhanwa Patra}$^{b,c}$\footnote[2]{sudhanwa@iitbhilai.ac.in},
{\bf Thomas Schwetz}$^a$\footnote[3]{schwetz@kit.edu},\\
{\bf Kiran Sharma}$^{a,b}$\footnote[4]{kirans@iitbhilai.ac.in}
\vspace{5mm}

$^a$ {\it {Institut f\"{u}r Astroteilchenphysik, Karlsruher Institut f\"{u}r Technologie (KIT), Hermann-von-Helmholtz-Platz 1, 76344 Eggenstein-Leopoldshafen, Germany}}\\
$^b$ {\it {Department of Physics, Indian Institute of Technology Bhilai, Durg-491002, India}}\\
$^c$ {\it
Institute of Physics, Sachivalaya Marg, Bhubaneswar- 751005, India}

\vspace{8mm} 

\today
  
\vspace{8mm} 

\abstract{Time reversal (T) symmetry violations in neutrino oscillations imply   the presence of an $L$-odd component in the transition probability at fixed neutrino energy, with $L$ denoting the distance between neutrino source and detector. Within the standard three-flavour framework, we show that the combination of the transition probabilities determined at the DUNE and T2HK experiments can establish the presence of an $L$-odd component, and therefore provide sensitivity to T violation, up to $4\sigma$ significance. The optimal neutrino energy window is from 0.68 to 0.92~GeV, and therefore a crucial role is played by the low-energy part of the DUNE event spectrum covering the second oscillation maximum. We compare the sensitivity to T violation based on this energy range using neutrino data only with the more traditional search for charge-parity (CP) violation based on the comparison of neutrino versus anti-neutrino beam data. We show that for DUNE it is advantageous to run in neutrino mode only, i.e., searching for T violating effects, whereas T2HK is more sensitive to CP violation, comparing neutrino and anti-neutrino data. Hence, the two experiments offer complementary methods to determine the complex phase in the PMNS mixing matrix.}

\end{center}

\end{titlepage}

\renewcommand{\thefootnote}{\arabic{footnote}}
\setcounter{footnote}{0}

\setcounter{page}{2}
\tableofcontents

\section{Introduction}

A genuine property of weak interactions with three fermion generations is the violation of the time reversal (T) or equivalently, the charge-parity (CP) conjugation symmetries of the fundamental theory \cite{Kobayashi:1973fv}.
This phenomenon may also take place in neutrino oscillations~\cite{Cabibbo:1977nk,Bilenky:1980cx,Barger:1980jm}, provided suitable non-trivial complex phases are present in the theory. 
Within the standard three-flavour paradigm, fundamental T and CP violation are equivalent and both are controlled by a single complex phase in the PMNS~\cite{Pontecorvo:1967fh,Maki:1962mu} mixing matrix, conventionally denoted as $\delCP$. Current data on neutrino oscillations \cite{T2K:2023smv,NOvA:2021nfi,Esteban:2024eli,Capozzi:2025wyn} show first indications for preferred regions for $\delCP$ and the search for effects of such complex phases is the main science goal of the next generation long-baseline oscillation experiments T2HK~\cite{Hyper-Kamiokande:2018ofw} and DUNE~\cite{DUNE:2020jqi,Abi:2020wmh}.

The usual strategy for this search is to perform a model-dependent fit to observed neutrino and anti-neutrino induced event spectra in terms of the three-flavour oscillation parameters $\Delta m^2_{31}$, $\Delta m^2_{21}$, $\theta_{12}$, $\theta_{13}$, $\theta_{23}$ as well as the complex phase $\delta_{\rm CP}$. This approach captures the full experimental information and offers the best sensitivity to determine the relevant parameters within the three-flavour framework, in particular $\delta_{\rm CP}$. The drawback of this approach is its model-dependence and the lack of intuitive physics observables related to the phenomenon of CP or T violation. In this work we address the second issue and provide an interpretation of the experimental results in terms of T violating observables (based on neutrino data only) contrasted to CP violating observables (from the comparison of neutrino and anti-neutrino data).

Traditionally, the T transformation in neutrino oscillations is related to the exchange of initial and final neutrino flavours: $\mathcal{T}[P_{\alpha\to\beta}(L)] = P_{\beta\to\alpha}(L)$ see e.g., refs.~\cite{Cabibbo:1977nk,Kuo:1987km,Krastev:1988yu,Toshev:1989vz,Toshev:1991ku,Arafune:1996bt,Parke:2000hu,Akhmedov:2001kd,Schwetz:2007py,Xing:2013uxa,Petcov:2018zka,Bernabeu:2019npc,Kitano:2024kdv,Bitter:2024eka,Yasuda:2024lwg} for an incomplete list of references. Due to the experimental difficulty to produce an electron or tau neutrino beam, this method cannot be pursued with conventional pion-based neutrino beams and probably requires the advent of a muon-based neutrino factory, see e.g.~\cite{Kitano:2024kdv} for a recent paper. Here, instead, we follow the approach of 
\cite{Schwetz:2021cuj,Chatterjee:2024jzt}, which is based on the well-known observation that the T transformation is equivalent to the formal replacement of the baseline $L\to -L$ but keeping the oscillation channel unchanged, i.e., $\mathcal{T}[P_{\alpha\to\beta}(L)] = P_{\alpha\to\beta}(-L)$ (see also the appendix of ref.~\cite{Chatterjee:2024jzt} for a discussion of this property). Hence, we can search for T violation by looking for an $L$-odd component of the transition probability $P_{\nu_\mu\to\nu_e}(L)$, considered as a function of $L$ but at fixed neutrino energy. 

In Refs.~\cite{Schwetz:2021cuj,Chatterjee:2024jzt} the emphasis has been put on model-independent measures for T violation. In the present paper we relax these ambitions with regard to model-independence and apply the approach of \cite{Schwetz:2021cuj} to the case of the standard unitary three-flavour mixing model. As we will show, this significantly increases the sensitivity, when the present knowledge \cite{Esteban:2024eli,Capozzi:2025wyn} about the mass and mixing parameters is taken into account. 
Technically, the analysis becomes very similar to the standard $\delCP$ determination. However, we argue that when results are interpreted in terms of the appearance probability as a function of baseline, we see explicitly the effect of T-odd contributions to the probability. We will study the potential of DUNE and T2HK to explore this effect, and contrast it to ``CP observables'' related to the comparison of neutrino versus anti-neutrino transition probabilities. 
As we will show, even with an appropriately chosen single energy bin with only neutrino beam running, DUNE has significant sensitivity to T violation. A crucial role in this respect will come from information from the second oscillation maximum, emphasizing the importance of the low energy part of the DUNE event spectrum, see e.g.~\cite{Ishitsuka:2005qi,Huber:2010dx,Coloma:2011pg,Qian:2013nhp,Wildner:2015yaa,Hyper-Kamiokande:2016srs,DeRomeri:2016qwo,Rout:2020emr} for previous studies on the 2nd oscillation maximum. In this sense, DUNE has better sensitivity to T violation (based on data only from neutrino beam mode), whereas T2HK preferably determines CP violation (based on the comparison of neutrino and anti-neutrino data).

The outline of the paper is as follows. In \cref{sec:formalism} we briefly introduce the oscillation formalism and give technical details on our numerical simulation of the DUNE and T2HK experiments. In \cref{sec:results} we discuss the numerical results about T violation from $L$ dependence of the oscillation probability: we identify the most sensitive energy range,  emphasizing the importance of the 2nd oscillation maximum in DUNE as well as the synergy with T2HK, and study the dependence of the sensitivity to prior knowledge on oscillation parameters as well as on the exposure. In \cref{sec:CP} we discuss the sensitivity to T violation compared to CP violation (based on the neutrino/anti-neutrino comparison) and work out the complementarity between the DUNE and T2HK experiments in this respect. In \cref{sec:XT} we consider the model-independent observable $X_T$ introduced in ref.~\cite{Chatterjee:2024jzt}, which is given by the difference of the oscillation probability at the baselines of the DUNE and T2HK experiments, evaluated at the same neutrino energy:
$X_T \equiv P_{\nu_\mu\to\nu_e}(L_2) - P_{\nu_\mu\to\nu_e}(L_1)$. Under certain conditions, a negative value of $X_T$ is a model-independent signal of T violation. Here we will investigate the sensitivity of DUNE and T2HK to the $X_T$ test within the standard three-flavour oscillation framework. We conclude in \cref{sec:conclusion}. In  \cref{sec:bins} we offer further details on the energy dependence of T and CP violation by studying the bin-per-bin sensitivity of the two experiments whereas in \cref{sec:density} we justify our approximation concerning the matter density.

\section{Formalism and analysis details}
\label{sec:formalism}

\subsection{Transition probabilities}
\label{sec:Ltest}

Let us briefly review the theoretical framework and fix our notation.
We focus on the $\nu_\mu\to\nu_e$ appearance probability $P_{\nu_\mu\to\nu_e}$, with the $L$ and $E_\nu$ dependence often left implicit. 
We approximate the matter density along the neutrino path being constant and the same for T2HK and DUNE. The impact of deviations from this approximation on the T-violation test has been quantified in \cite{Schwetz:2021thj} and found to be small, see also \cite{Bitter:2024eka}. Then the oscillation probability is obtained as
\begin{equation}\label{eq:P}
\begin{split}    
  P_{\nu_\mu\to\nu_e} &= \left| \sum_{i=1}^3 c_i e^{-i\lambda_iL}\right|^2  
    \,, \qquad c_i \equiv U_{\mu i}^{m*} U_{e i}^m \,,
\end{split}
\end{equation}
where $\lambda_i$ are the eigenvalues of the effective Hamiltonian including the matter potential \cite{Wolfenstein:1977ue}, and $U_{\alpha i}^m$ is the effective mixing matrix in matter. Because $U_{\alpha i}^m$ is unitary, only two out of the three $c_i$ coefficients are independent, which we take as $c_2,c_3$.
One can split the probability into T-even and T-odd parts:
\begin{align}
  P_{\nu_\mu\to\nu_e} = P_{\rm even} + P_{\rm odd} \,,   
\end{align}
with
\begin{align}
  P_{\rm even} &= 
    4|c_2|^2\sin^2\phi_{21}
  + 4|c_3|^2\sin^2\phi_{31} 
  + 8{\rm Re}[c_2^*c_3]\sin\phi_{21}\sin\phi_{31}\cos\phi_{32}
  \label{eq:Peven}\\[2mm]
  P_{\rm odd} &= 
  8{\rm Im}[c_2^*c_3]\sin\phi_{21}\sin\phi_{31}\sin\phi_{32}
  \,, \label{eq:Podd}
\end{align}
where
\begin{align}\label{eq:phi}
  \phi_{ij} \equiv \frac{\lambda_i-\lambda_j}{2}L = 
  \frac{\Delta m^2_{ij,{\rm eff}}(E_\nu) L}{4E_\nu} \,,
\end{align}
with $\Delta m^2_{ij,{\rm eff}}(E_\nu)$ denoting the effective mass-squared differences in matter. Note that in \cref{eq:Peven,eq:Podd} we see explicitly the even and odd behaviour under $L\to -L$, respectively. The factor ${\rm Im}[c_2^*c_3] = {\rm Im}[   U_{\mu 2}^m U_{e 2}^{m*}  U_{\mu 3}^{m*} U_{e 3}^m]$ in \cref{eq:Podd} corresponds to the Jarlskog invariant \cite{Jarlskog:1985ht} for mixing in matter, see also \cite{Freund:2001pn,Denton:2019yiw} for discussions.

Considering now the transition probability as a function of baseline for fixed $E_\nu$, T violation can be established if data cannot be fitted with $P_{\rm even}(L)$ alone. 
In Ref.~\cite{Schwetz:2021cuj}, this test has been formulated in a general way, imposing no prior knowledge on the coefficients $c_i$ in order to remain as model-independent as possible. In contrast, in the following we will specialize to the standard three-flavour scenario by using the specific expressions for $c_i$ and $\phi_{ij}$ implied by \cref{eq:P,eq:phi}, and impose the prior knowledge on the three mixing angles $\theta_{ij}$, mass-squared differences $\Delta m^2_{ij}$, as well as the Standard Model matter potential. In this way, the search for T violation can be applied even with a single measurement at suitable chosen baseline and energy. 

In order to obtain the CP conjugated anti-neutrino transition probability $P_{\bar\nu_\mu\to\bar\nu_e}$, \emph{two} modifications are necessary in \cref{eq:P}: complex conjugation has to be applied to the $c_i$ coefficients and the sign of the matter potential has to be changed, the latter leading to modification of the absolute values of $c_i$ as well as of the eigenvalues $\lambda_i$. We see that in the presence of matter, a decomposition into CP-even and CP-odd components is less straight-forward than in the case of the T transformation and offers only an indirect interpretation in terms of fundamental CP violation of the theory because of environmental CP violation induced by background matter.  

\subsection{Analysis details} 
\label{sec:simulation}

In this subsection, we briefly describe the experimental specifications of the two long-baseline experiments DUNE and T2HK assumed in this work and the adopted statistical analysis.

\bigskip

\noindent {\bf DUNE}: The Deep Underground Neutrino Experiment (DUNE)  will generate an intense beam of neutrinos at Fermi National Accelerator Laboratory (Fermilab) in Illinois and direct it toward the far detector located approximately 1300~km away at the Sanford Underground Research Facility (SURF) in South Dakota.  We have adopted the experimental setup described in the Technical Design Report (TDR) \cite{DUNE:2020jqi,DUNE:2021cuw}, including systematic uncertainties and detector efficiencies.  The TDR configuration consists of a far detector made of a 40~kton Liquid Argon Time Projection Chamber, offering exceptional tracking and particle identification capabilities. The current plan follows a staged deployment with up to four 17~kton detector modules. DUNE will use a high-intensity 120~GeV proton beam with a power of 1.2~MW, expected to deliver $1.1\times 10^{21}$ protons on target (P.O.T) per year. For our analysis, the low-energy tail of the event spectrum in DUNE is relevant, which suffers from low event numbers, see \cite{Chatterjee:2024jzt} for a discussion. Therefore, results are largely limited by DUNE statistical uncertainties. As our default exposure we assume 336~kt~MW~yr, which should be achieved roughly after 7 years of operation \cite{IRN-talk}. As we focus on the sensitivity for neutrinos, we assume that the full exposure is taken in the neutrino beam mode. Anti-neutrino running is considered in \cref{sec:CP} and \cref{sec:bins}.
For the energy resolution we assume an improved  performance, as proposed in Refs.~\cite{DeRomeri:2016qwo,Friedland:2018vry} (see also \cite{Kopp:2024lch,Chatterjee:2021wac}). We consider a Gaussian detector resolution with $\sigma= \alpha E_\nu + \beta \sqrt{E_\nu} + \gamma$, where $E_\nu$ is the neutrino energy in GeV. Our default assumption is 
$(\alpha,\beta,\gamma)= (0.045,0.001,0.048)$,
with units in GeV. This choice corresponds to the resolution from \cite{Friedland:2018vry,Chatterjee:2021wac}, see also \cite{Chatterjee:2024jzt} for a discussion.

\bigskip

\noindent{\bf T2HK}: The Tokai-to-hyper-Kamiokande (T2HK) experiment is a next-generation off-axis accelerator-based neutrino project designed to study neutrino oscillations with a 295~km baseline. It uses the existing 30~GeV proton beam from the J-PARC accelerator facility in Tokai, Japan, to produce intense beams of neutrinos and antineutrinos. The experiment will employ a massive Water Cherenkov far detector located at the Hyper-Kamiokande (HK) site, with a fiducial volume of 187 kilotons. To assess the physics reach of T2HK, we follow the experimental specifications detailed in the Hyper-Kamiokande design report \cite{Hyper-Kamiokande:2018ofw}. The total integrated exposure is expected to be $1.3~ \rm{MW}\times 10\times 10^7$ seconds, corresponding to $2.7\times 10^{22}$ protons on the target (PT). Our default exposure for T2HK it is 608~kt~MW~yr, corresponding roughly to 2.5~yr of running in the neutrino mode. As the relevant energy window for the T-violation test is located close to the maximum of the event spectrum, T2HK statistical errors are subdominant. In our analysis, we adopt a simplified treatment of systematic uncertainties, including an uncorrelated 5\% (3.5\%) signal normalization error, a 10\% background normalization error, and a 5\% energy calibration error. 
For the energy resolution we use the same parameterization as for DUNE as mentioned above, with the values 
$(\alpha,\beta,\gamma)=
 (0.12,0.07,0.0)$, which match the resolution provided  in the design report~\cite{Hyper-Kamiokande:2018ofw}.

\bigskip

In order to estimate the sensitivity to T-violation in neutrino oscillations, we use the GLoBES software package~\cite{Huber:2004ka,Huber:2007ji}, implementing necessary modifications to accommodate our specific analysis framework. Our study focuses exclusively on the appearance channel, as this channel provides the most direct probe of T-asymmetries in the neutrino sector. In our analysis, we treat the mass-squared splittings $\Delta m^2_{21}$ and $\Delta m^2_{31}$ as fixed parameters, constrained by high-precision external measurements, such as the disappearance channel from DUNE/T2HK or the JUNO reactor experiment \cite{JUNO:2022mxj}. The neutrino mass ordering is assumed to be known and fixed to normal ordering. Consequently, the only free parameters in our oscillation fit are the three mixing angles $\theta_{12}$, $\theta_{13}$ and $\theta_{23}$. 

The T-violation test is performed by comparing the transition probabilities at fixed neutrino energies. Therefore, we consider identical reconstructed energy bins for both T2HK and DUNE of bin width 0.12~GeV. The T2HK beam is rather narrow and centred roughly between 0.2 and 1.1~GeV, whereas the DUNE beam has a boad peak between 2 and 4~GeV, but the low energy tail extends down to approximately 0.7~GeV. This low energy tail will play a crucial role in our analysis, as it covers the 2nd oscillation maximum and offers an overlap region with T2HK. Below we will give some discussion of the T and CP sensitivity of various energy bins. 

For the T-violation analysis we only use simulated data in the neutrino beam mode. In principle, an analogous analysis can be performed also with anti-neutrinos. However, due to the sizeable intrinsic neutrino component of the anti-neutrino beams both for T2HK and DUNE, it is not possible to extract the anti-neutrino transition probability in a clean way from the data. Hence, data in the anti-neutrino beam mode does not allow a straight-forward interpretation in terms of $L$-odd versus $L$-even components and T-violating observables are significantly diluted \cite{Chatterjee:2024jzt}. Therefore, we restrict the T violation analysis to the neutrino mode only, where the intrinsic anti-neutrino contamination is sufficiently small to be neglected. 

\begin{table}[t]
  \centering
  \begin{tabular}{l@{\quad}c@{\quad}c@{\quad}c@{\quad}c@{\quad}c}
    \hline\hline
    parameter: & $\sin^2{\theta_{12}}$ & $\sin^2{\theta_{13}}$ & $\sin^2{\theta_{23}}$ & $\Delta m^2_{21}$ [eV$^2$]& $\Delta m^2_{31}$  [eV$^2$] \\
    value: &
    0.307 & 0.022 & 0.561 & $7.49\times 10^{-5}$ & $2.53\times 10^{-3}$ \\
$1\sigma$ uncert.: &
 $3.9\%$ & $2.54\%$ & $2.32\%$ & fixed & fixed \\
    \hline\hline
  \end{tabular}
  \mycaption{Best-fit values and $1\sigma$ uncertainties of the standard three-flavour neutrino parameters \cite{Esteban:2024eli} adopted in the numerical analysis.}
\label{tab:osc-params}
\end{table}

We use the following $\chi^2$-function:
\begin{equation}
\label{eq:chi2k}
  \chi^2_{x,i}(\theta) = \min\limits_{\xi_x}\left[
    2\left(N_i^x(\theta,\xi_x) -N_i^{x,\rm obs} - N_i^{x,\rm obs} \ln \frac{N_i^x(\theta,\xi_x)}{N_i^{x,\rm obs}}\right) + \sum_{\xi_x} \frac{\xi_x^2}{\sigma_\xi^2} \right] 
    + \chi^2_{\rm prior}(\theta)
    \,.
\end{equation}
Here $\theta$ collectively denotes the standard three-flavour mixing angles $\theta_{ij}$, and $N_i^x(\theta,\xi_x)$ is the number of events predicted in energy bin $i$ for experiment $x={\rm T2HK, DUNE}$ assuming T conservation, i.e., $\delta_{\rm CP} = 0,\pi$, calculated including backgrounds and various systematics as described above. The latter are presented by pull parameters, generically denoted by $\xi_x$ in \cref{eq:chi2k}.
$N_i^{x,\rm obs}$ is the corresponding ``observed'' number of events, which will depend on the true mechanism of neutrino conversion realised in Nature. Unless stated otherwise, we will assume the benchmark values of the standard three-flavour oscillation parameters shown in \cref{tab:osc-params} to calculate $N_i^{x,\rm obs}$. Then we study the sensitivity to T-violation as a function of the assumed true value of $\delta_{\rm CP}$. 
We use a line-averaged constant matter density of 2.84~g/$\rm{cm}^3$~\cite{stacey:1977,PREM:1981} for both, DUNE and T2HK, which is a good approximation for these baselines \cite{Schwetz:2021thj}. We provide a quantitative estimate of this approximation in \cref{sec:density}.
The last term in \cref{eq:chi2k} takes into account external knowledge on the three mixing angle parameters. We assume a Gaussian prior on $\sin^2\theta_{ij}$, centred at the values given in \cref{tab:osc-params}. For the standard deviation we adopt benchmark values based on the current uncertainties, see \cref{tab:osc-params}. We will show, how the sensitivity to T violation depends on the assumed prior, including also assumptions about the octant degeneracy of $\theta_{23}$.

Finally we define,
\begin{equation}\label{eq:chi2T}
  \Delta \chi^2 = \min\limits_{\theta;\delta_{\rm CP}^{\rm fit}=0,\pi}\left[\sum_{x,i}\chi^2_{x,i}(\theta)\right] \,,
\end{equation}
and we interpret our sensitivity to T violation by evaluating $\Delta \chi^2$ for 1 degree of freedom, where $\sqrt{\Delta \chi^2}$ quantifies the significance of excluding the T conservation hypothesis (under the Gaussian approximation). Depending on the analysis, we sum or do not sum over energy bins or over the two experiments. When studying CP violation, the sum over $x$ includes also neutrino and anti-neutrino runs for both experiments.

\section{T violation from $L$ dependence}
\label{sec:results}

\begin{figure}[t!]
\centering
\includegraphics[width=0.32\textwidth]{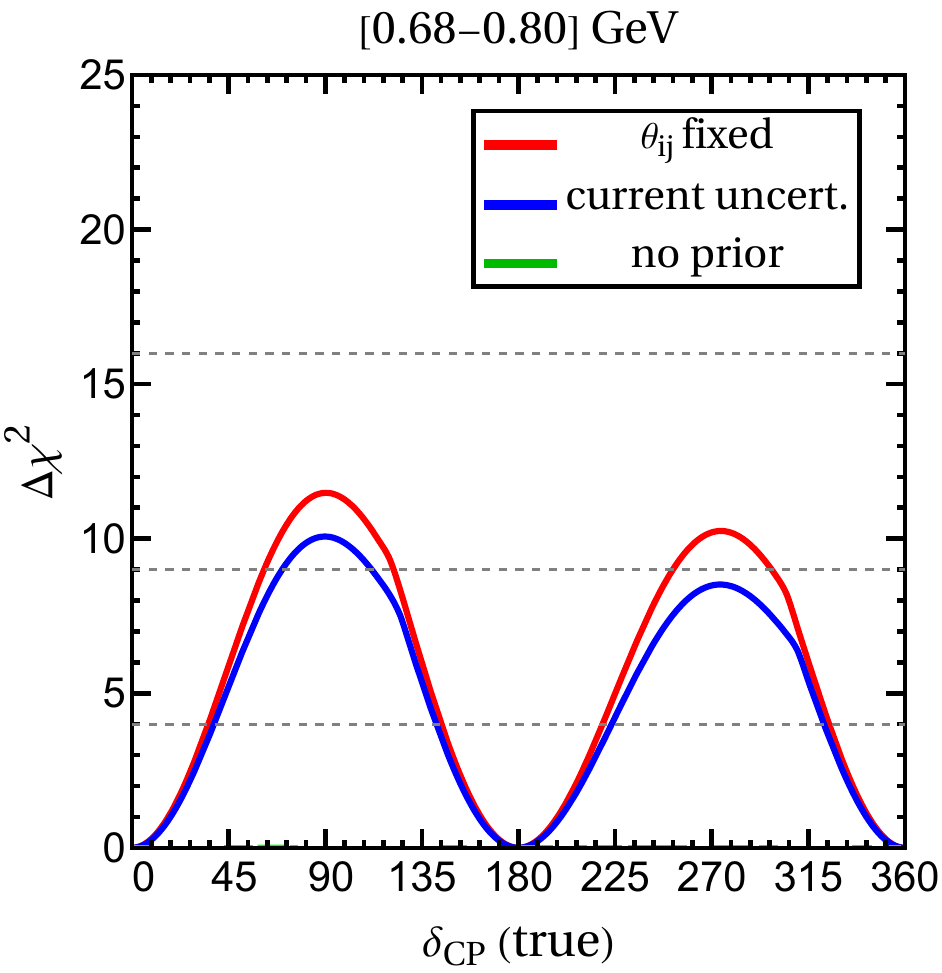}
\includegraphics[width=0.32\textwidth]{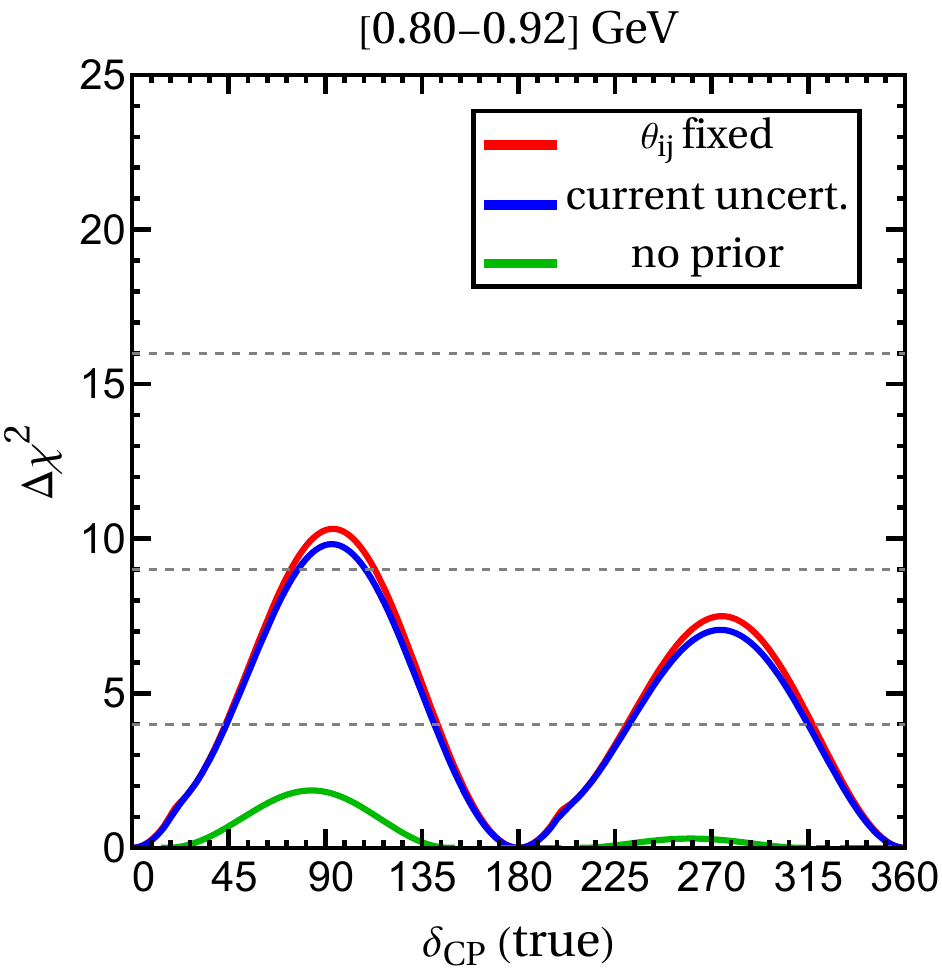}
\includegraphics[width=0.32\textwidth]{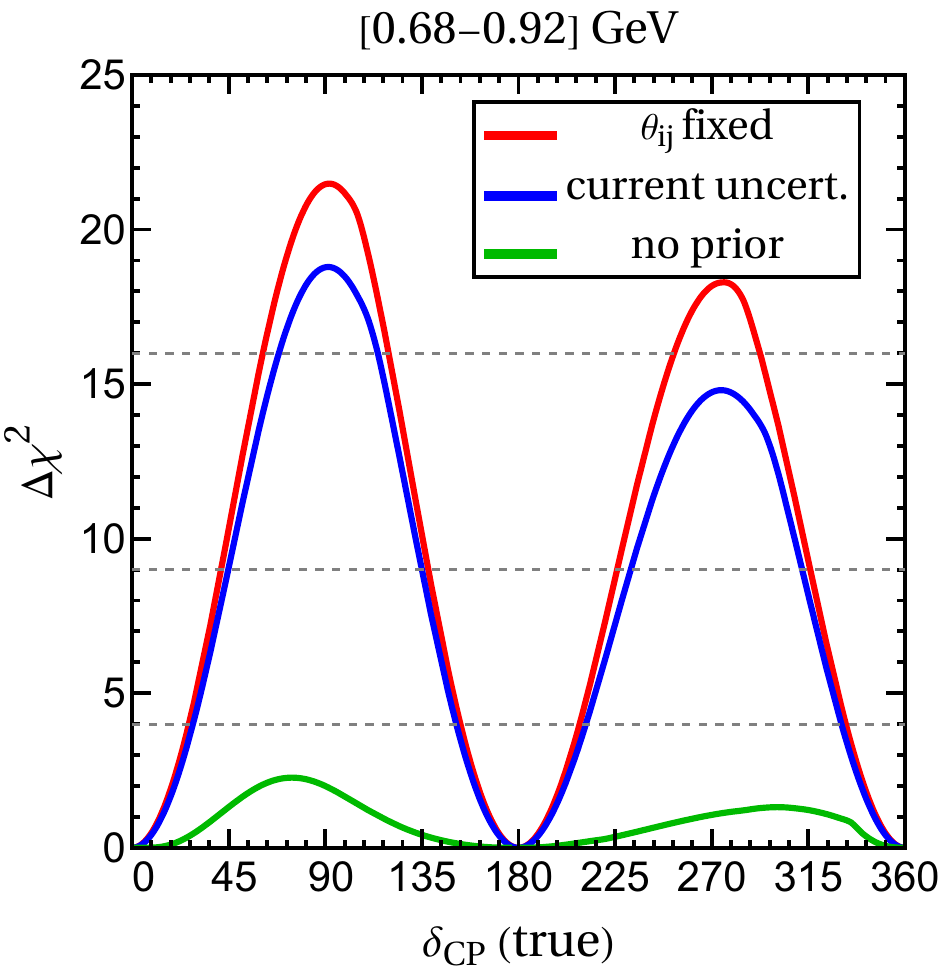}
\mycaption{Sensitivity to T violation for DUNE+T2HK as a function of the true value of $\delta_{\rm CP}$ for two  energy bins between 0.68 and 0.92~GeV. Left and middle panels show the bins separately and the right panel corresponds to the combination. Different curves correspond to different assumptions on prior knowledge on the mixing angles $\theta_{ij}$, namely no prior knowledge (green), uncertainties corresponding to current global fit uncertainties (blue), and mixing angles fixed to their best fit points (red).}
\label{fig:chi2-delta}
\end{figure}

\Cref{fig:chi2-delta} shows the sensitivity to T violation as a function of the true value of $\delta_{\rm CP}$ for our default configuration described above. We show the sensitivity for two energy bins $[0.68,0.8]$~GeV and $[0.80,0.92]$~GeV separately (left and middle panels) as well as combined (right panel). We use neutrino beam mode only; therefore, the sensitivity to $\delta_{\rm CP}$ can only emerge from the presence of an $L$-odd component in the probability. The right panel shows excellent sensitivity around $4\sigma$ in case of close-to-maximal T violation.

\begin{figure}[t!]
\centering
\includegraphics[width=0.32\textwidth]{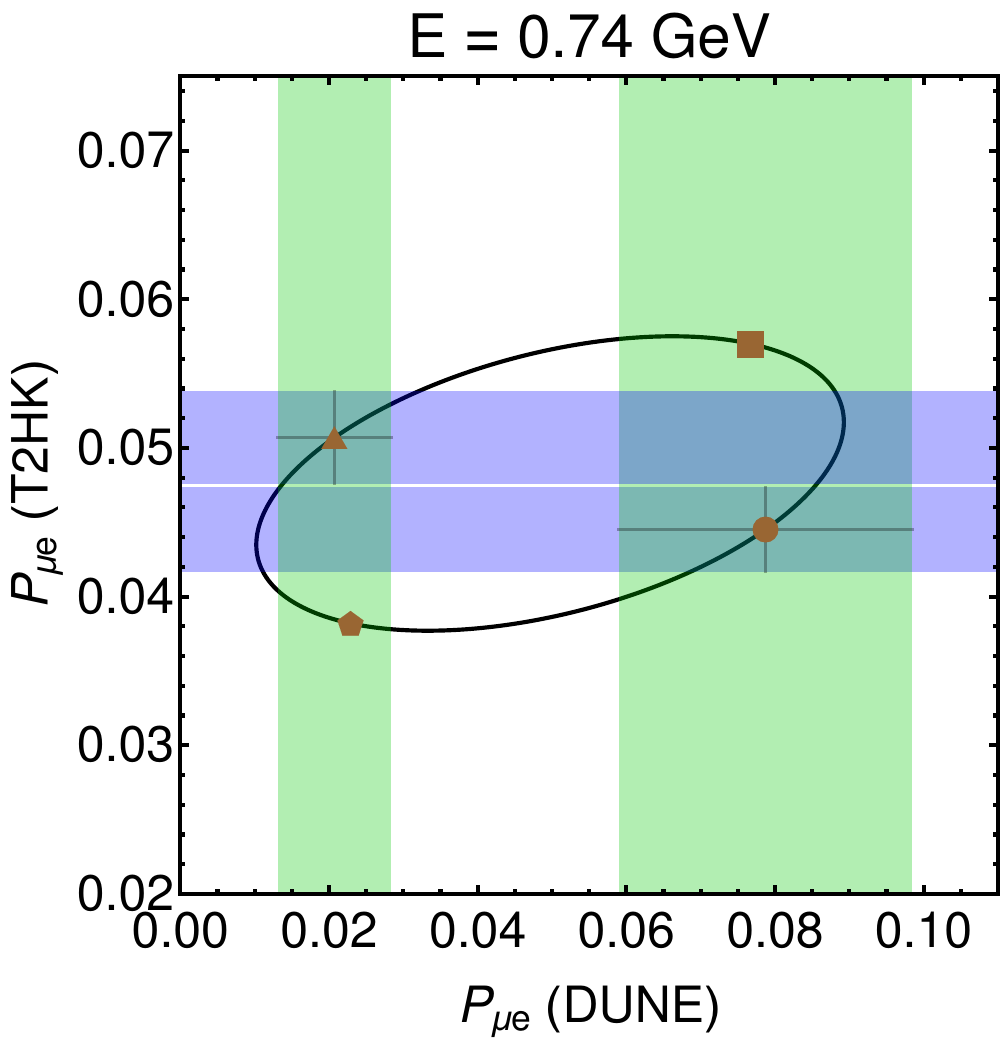}
\includegraphics[width=0.32\textwidth]{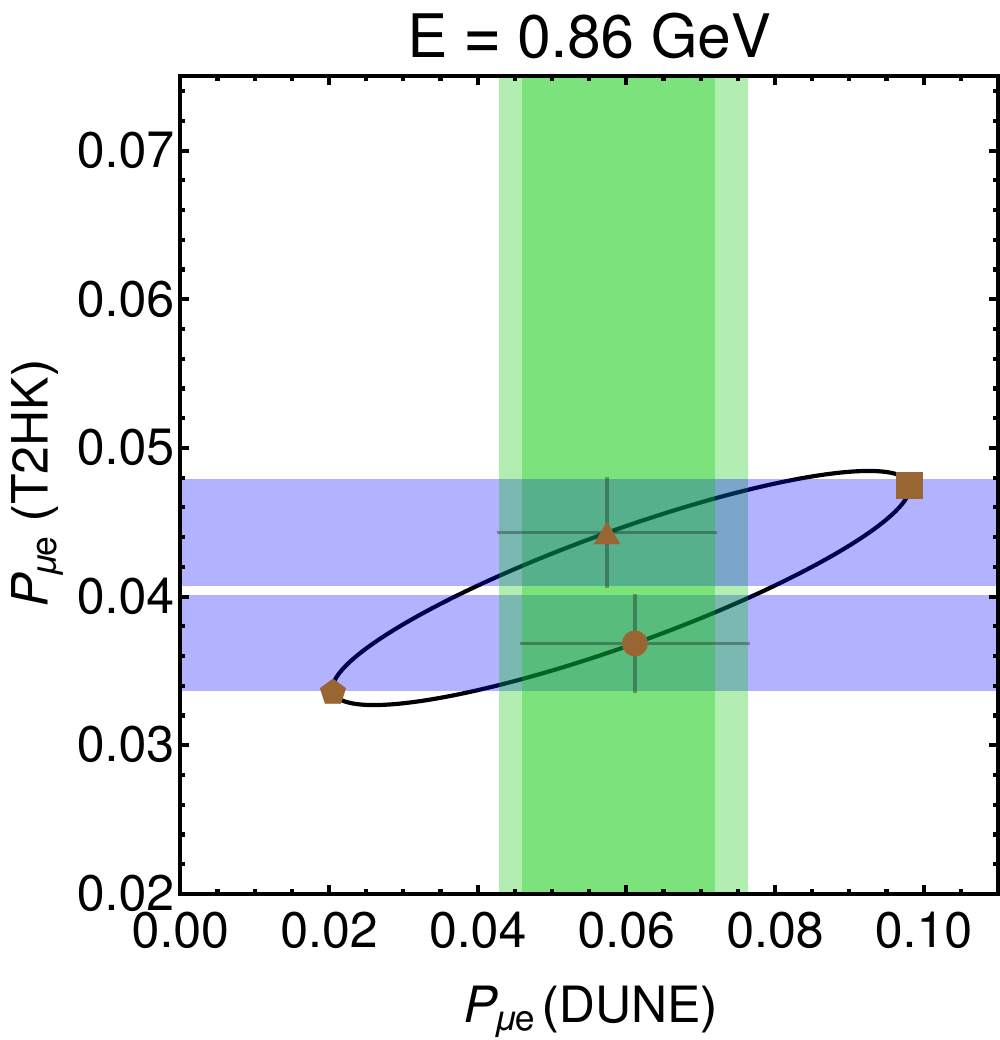}
\includegraphics[width=0.32\textwidth]{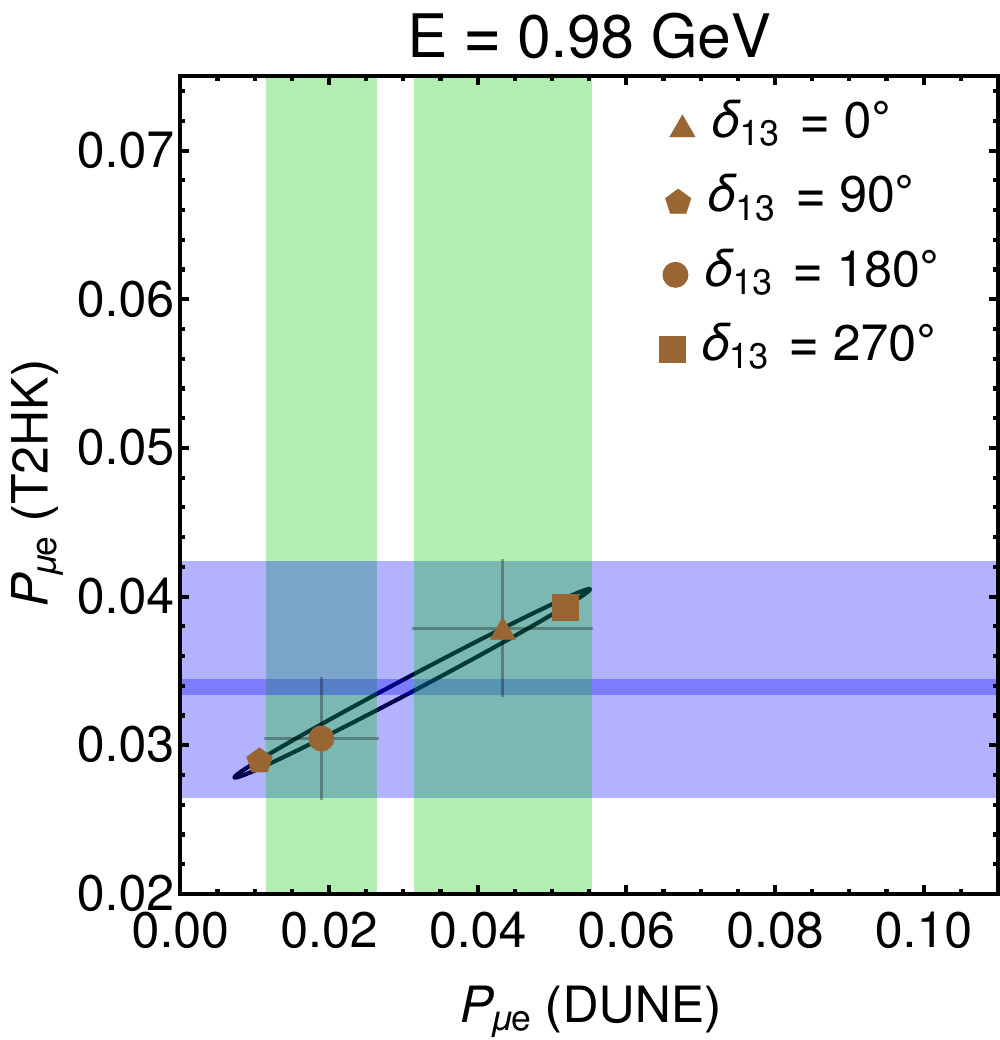}
\mycaption{Bi-probability plot for $P_{\nu_\mu\to\nu_e}$ for DUNE and T2HK for three energies where there is overlap between the two experiments. Oscillation parameters are chosen according to \cref{tab:osc-params} and the symbols indicate values of $\delta_{\rm CP}$. We indicate the $\pm1\sigma$ error bands for the T-conserving cases, using the expected number of events in a bin with width 0.12~GeV centred at the energies of each panel, for our default assumptions about exposure.}
\label{fig:bi-prob}
\end{figure}

We have found that the two energy bins considered in \cref{fig:chi2-delta} provide optimal sensitivity to T violation, although it is interesting to remark that the sensitivity has different origin for the two bins: 
for the energy bin $[0.80,0.92]$~GeV, DUNE dominates the sensitivity and T2HK gives only a sub-leading contribution. 
This becomes clear from the bi-probability plots in \cref{fig:bi-prob}, in the plane of $(P_{\nu_\mu\to\nu_e}^{\rm DUNE},P_{\nu_\mu\to\nu_e}^{\rm T2HK})$. The middle panel corresponds to the central energy of the $[0.80,0.92]$~GeV bin. By comparing the location of the maximally T-violating values of $\delta_{\rm CP}$ at the outer edges of the probability ellipse with the error bands corresponding to T conservation, we see that DUNE has excellent sensitivity to exclude these points, whereas for T2HK T conservation and maximal T violation are consistent within $1\sigma$. 
In contrast, for the energy bin $[0.68,0.8]$~GeV, the good sensitivity shown in the left panel of \cref{fig:chi2-delta} emerges from a synergy between the two experiments. As visible in the left panel of \cref{fig:bi-prob}, the individual error bands for T conservation (shown by green and blue shaded bands) are close to the points corresponding to $\delCP = 90^\circ$ and $270^\circ$, whereas the combination of the two experiments, i.e., the overlap regions of the bands marked with the error bars, are located at sizeable distances from the maximally T-violating points. The right panel in \cref{fig:bi-prob} illustrates the absence of sensitivity for the next higher energy bin.

Further details on bin-wise and individual sensitivities are given in \cref{sec:bins}. Let us emphasize the crucial importance of the low-energy tail of the DUNE event spectrum for this analysis, covering the 2nd oscillation maximum.
It is well known that the second oscillation maximum provides high sensitivity to $\delCP$, see, for instance,\cite{Rout:2020emr,Wildner:2015yaa,Qian:2013nhp,Coloma:2011pg,Ishitsuka:2005qi,Schwetz:2021cuj}. Here we argue, that this is a manifestation of T violation, as no neutrino/anti-neutrino comparison is involved but only the $L$ dependence of the oscillation probability (see further discussions in \cref{sec:CP} and \cref{sec:bins}). 

\bigskip

\Cref{fig:chi2-delta} illustrates also the dramatic impact of prior knowledge of the oscillation parameters on the sensitivity. For the green curves, no prior information on the mixing angles is assumed and we obtain a marginal sensitivity well below $2\sigma$. This situation is similar (though not identical) to the model-independent approach of our previous paper \cite{Chatterjee:2024jzt}.\footnote{Note that in \cite{Chatterjee:2024jzt} we have assumed a DUNE exposure of 840~kt~MW~yr in order to boost the sensitivity, compared to the default exposure of 336~kt~MW~yr assumed here.} In contrast, if a prior corresponding to the uncertainties from the current global fit on $\sin^2\theta_{ij}$ is imposed (blue curves in \cref{fig:chi2-delta}), the sensitivity increases drastically, reaching more than $3\sigma$ for $\delCP \approx 90^\circ$ for individual bins and more than $4\sigma$ for the combination. We note that at the time of the DUNE and T2HK experiments, an even better accuracy would be available from JUNO on $\theta_{12}$ and $\theta_{13}$ \cite{JUNO:2022mxj} and from the DUNE/T2HK disappearance channels on $\theta_{23}$. Comparison of the blue and red curves in \cref{fig:chi2-delta} shows that a perfect knowledge of the mixing angles leads only to a slight improvement compared to present uncertainties.

\begin{figure}[t!]
\centering
\includegraphics[width=0.48\textwidth]{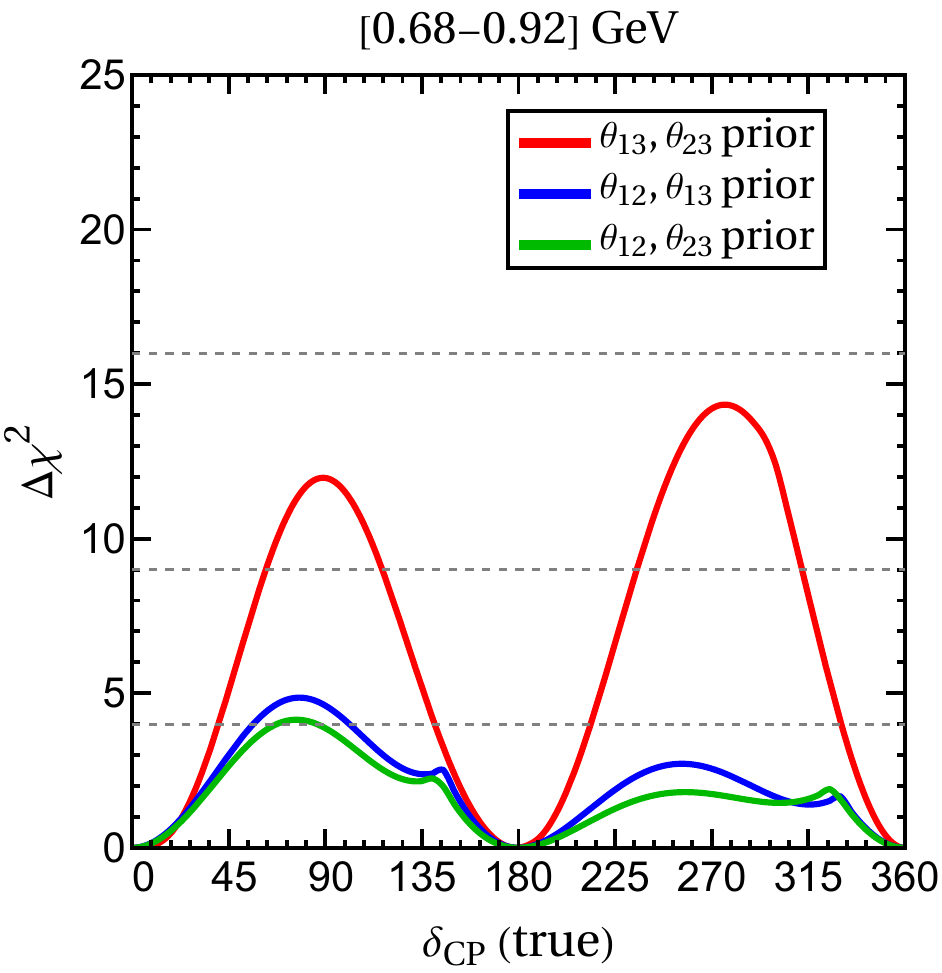}
\includegraphics[width=.48\textwidth]{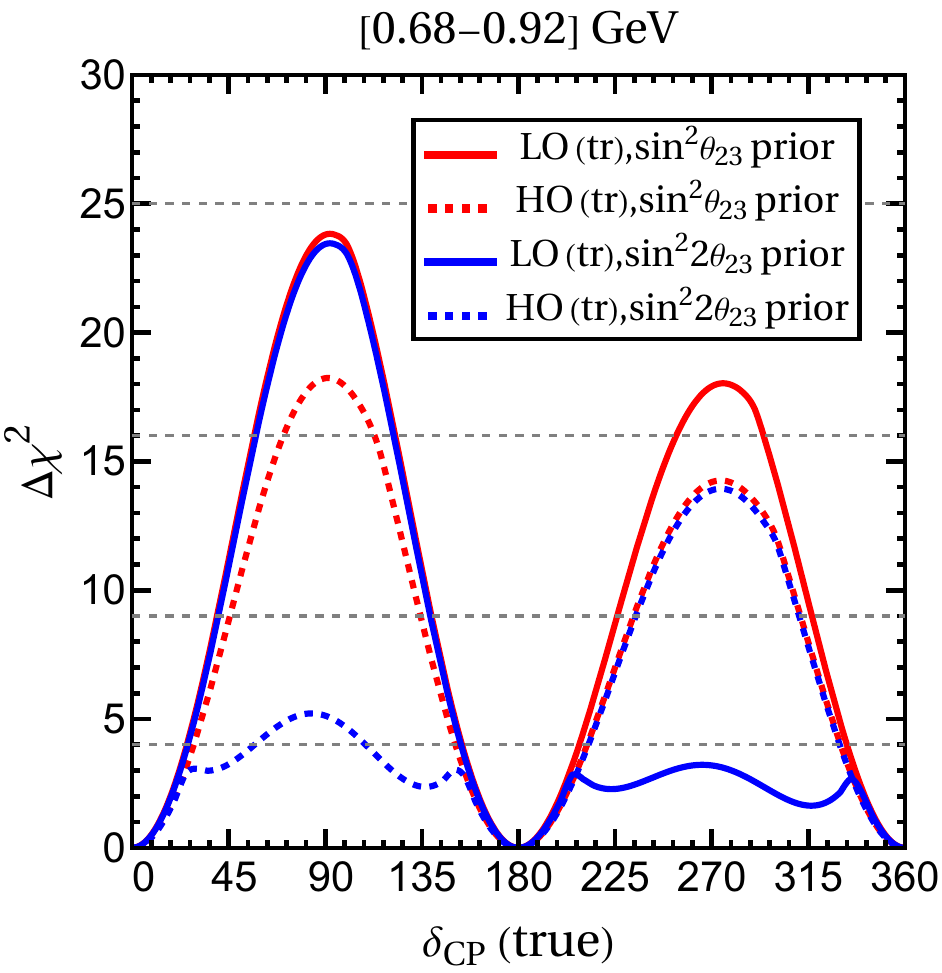}
\mycaption{Dependence of the sensitivity to T violation on prior knowledge on mixing angles. The left panel assumes a prior corresponding to current global fit uncertainties for two mixing angles and no prior for the third one. The right panel shows the dependence on the $\theta_{23}$ octant: for the solid (dashed) curves we assume a true value of $\theta_{23}$ in the 1st (2nd) octant, namely 
 $\sin^2\theta_{23} = 0.435\, (0.565)$; for the red (blue) curves we adopt a prior on $\sin^2\theta_{23}$ ($\sin^22\theta_{23}$), with an uncertainty of $\sigma_{\sin^2\theta_{23}} = 0.015$ ($\sigma_{\sin^22\theta_{23}} = 0.01$). }
\label{fig:octant}
\end{figure}

The left panel of \cref{fig:octant} shows that indeed the knowledge of all three mixing angles is important to reach such good sensitivities as displayed in \cref{fig:chi2-delta}, especially on $\theta_{13}$ and $\theta_{23}$.
Furthermore, an important effect comes also from the $\theta_{23}$ octant degeneracy. In our default analysis we assume a best fit value in the second octant, $\sin^2\theta_{23} = 0.561$, and impose a prior on $\sin^2\theta_{23}$, which does determine the octant and therefore breaks the well-known octant degeneracy \cite{Fogli:1996pv}. In the right panel of \cref{fig:octant} we study this issue in some detail. For the red curves we assume a prior on $\sin^2\theta$ (breaking the degeneracy), whereas for the blue curves the prior is imposed on $\sin^22\theta$, which is symmetric with respect to the octant. By comparing the red and blue curves we see that the knowledge of the $\theta_{23}$ octant is rather important, in particular for the two combinations of true values ($\theta_{23}>45^\circ$, $\delta_{\rm CP} \simeq 90^\circ$) and ($\theta_{23}<45^\circ$, $\delta_{\rm CP} \simeq 270^\circ$). If the octant is known, the sensitivity is somewhat better for $\theta_{23}^{\rm true} < 45^\circ$. While \cref{fig:octant} emphasizes the important to resolve the octant degeneracy, we note that the experiments considered here have good sensitivity to indeed lift this degeneracy, see e.g.~\cite{Ballett:2016daj,Chatterjee:2017irl,DUNE:2020jqi,Agarwalla:2024kti}. Therefore, we adopt the prior on $\sin^2\theta_{23}$ as our default assumption, assuming that the octant is known by the time the data becomes available. 

\begin{figure}[t!]
\centering
\includegraphics[width=0.49\textwidth]{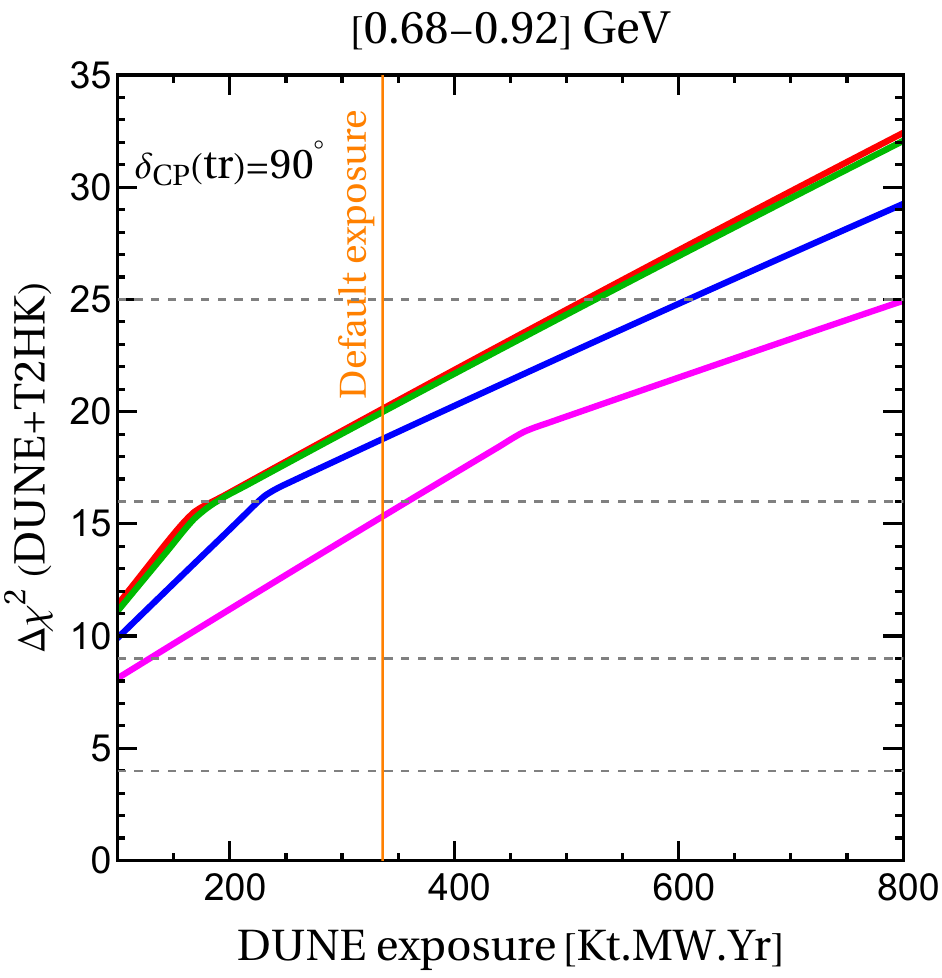}
\includegraphics[width=0.49\textwidth]{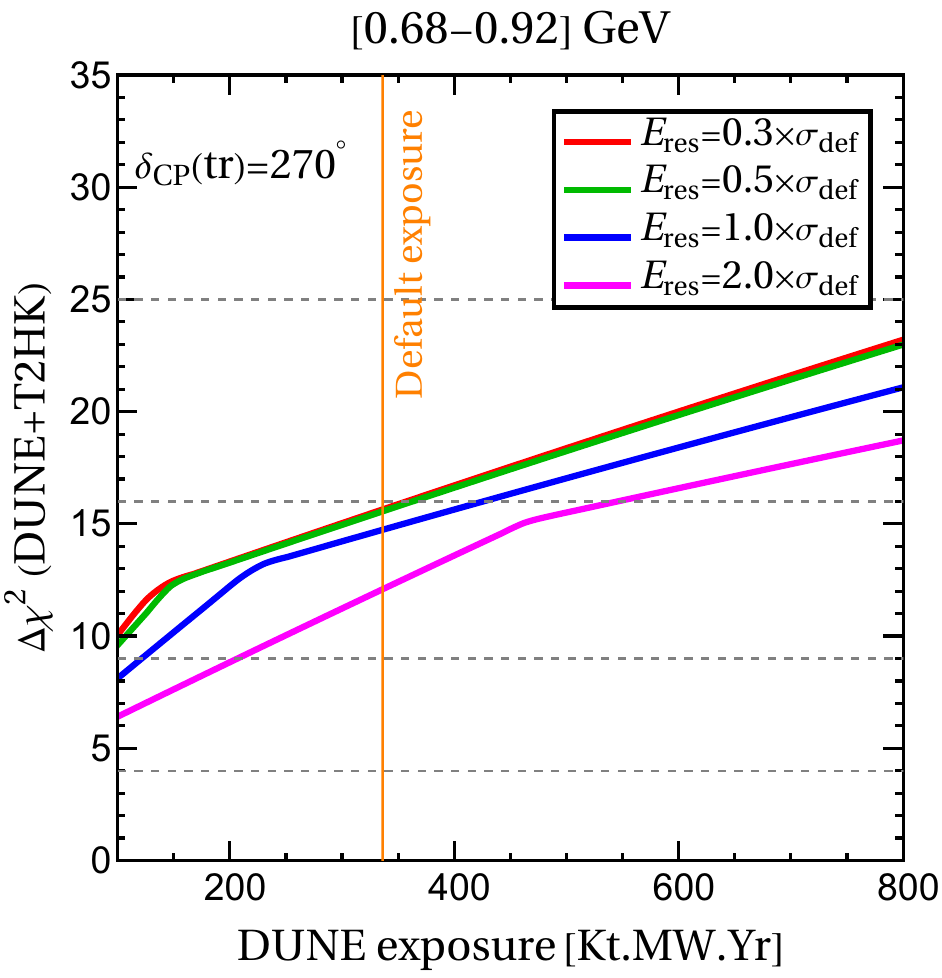}
\mycaption{Sensitivity to T violation as a function of the DUNE exposure in the neutrino beam mode for $\delta_{\rm CP}^{\rm true} = 90^\circ \, (270^\circ)$ for the left (right) panel. The T2HK exposure is kept fixed at 608~kt~MW~yr. For reference, our default DUNE exposure of 336~kt~MW~yr is obtained roughly after 7~years of operation. We show sensitivities from combining the two energy bins $[0.68,0.8]$ and $[0.8,0.92]$~GeV. Different curves correspond to re-scaling the energy resolution of DUNE by factors of 2, 1, 0.5, 0.3 compared to the default assumption stated in \cref{sec:simulation} which corresponds to the blue curves.}
\label{fig:chi2-exposure}
\end{figure}

\bigskip

As the considered energy range is in the low-energy tail for DUNE, small event numbers are expected. Therefore, we study the sensitivity as a function of the DUNE exposure in \cref{fig:chi2-exposure}, together with the assumed energy resolution. We have checked that the results depend only mildly on both, the exposure and energy resolution in T2HK. For our default assumption on the energy resolution \cite{Friedland:2018vry} as quoted in \cref{sec:simulation} (blue curve), sensitivities above $4\sigma$ are obtained for exposures above 250 (400)~kt~MW~yr in the neutrino running mode for a true $\delCP=90^\circ$ ($270^\circ$). Relaxing the requirements on energy resolution by a factor 2 (magenta curve) still allows for sensitivities above $3\sigma$ for exposures above 200~kt~MW~yr. For energy resolutions even better than our default assumption correspondingly smaller exposures are needed. We observe that the sensitivity saturates by resolutions better than a factor 2 compared to our default values.

The kink in the sensitivity curves in \cref{fig:chi2-exposure} is related to the lower energy bin $[0.68,0.8]$~GeV. As discussed above, there the sensitivity emerges from an interplay of DUNE and T2HK. As visible in the left panel of \cref{fig:bi-prob}, we expect that when decreasing the statistical error band for DUNE, the best fit jumps between $\delCP=0$ and $\pi$. For small enough DUNE errors, the sensitivity is determined by T2HK error bars which are kept fixed in \cref{fig:chi2-exposure}, and therefore explaining the reduced slope as a function of DUNE exposure.

\section{T versus CP}
\label{sec:CP}

In this section we compare the sensitivity to T violation based on the $L$ dependence for neutrino data only with the more traditional way of considering event spectra for neutrinos and anti-neutrinos as motivated by the CP symmetry. 
Hence, contrary to our default analysis described in the previous sections, we include now also simulated data with an anti-neutrino enhanced beam. With a slight abuse of language, we refer in the following to ``T-violation'' when neutrino data only is used, and to ``CP-violation'' when neutrino and anti-neutrino data are combined. Let us mention, however, that when neutrino and anti-neutrino event numbers are fitted together the ``T-violation'' information is automatically included as well, modulo reduced statistics when the total exposure is shared between neutrino and anti-neutrino modes.

\begin{figure}[t!]
    \centering
    \includegraphics[width=0.49\linewidth]{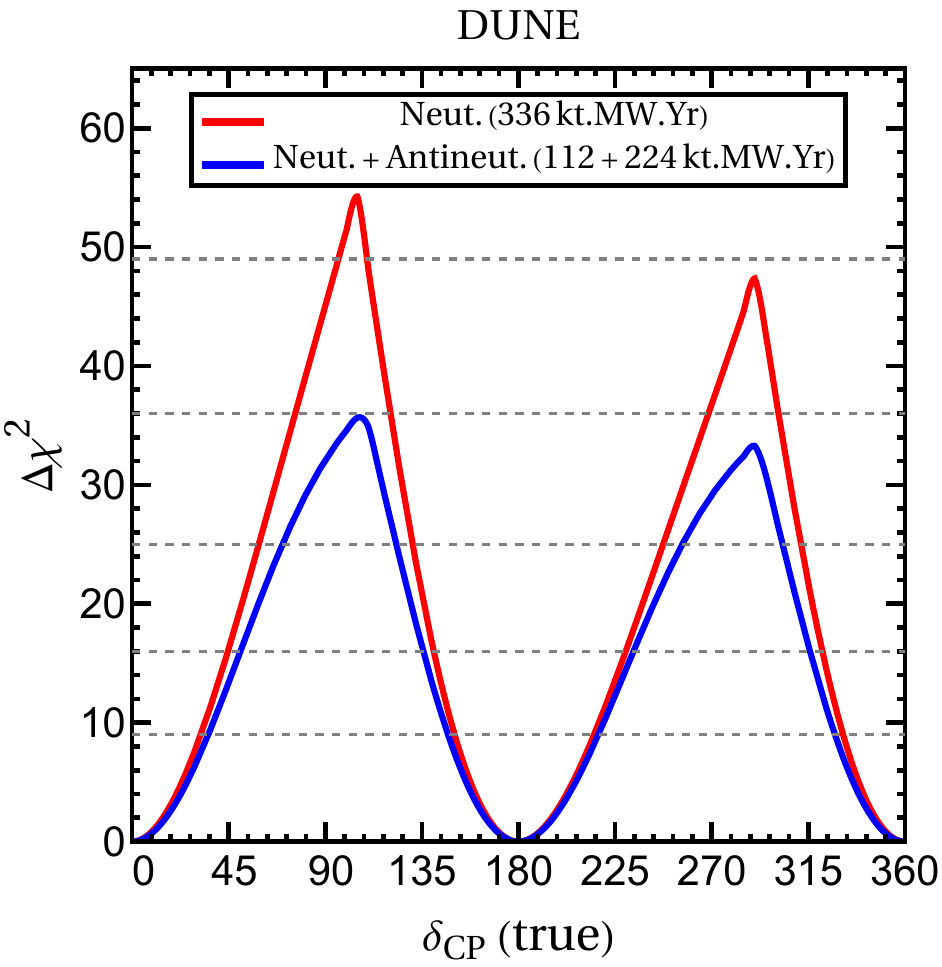}
     \includegraphics[width=0.49\linewidth]{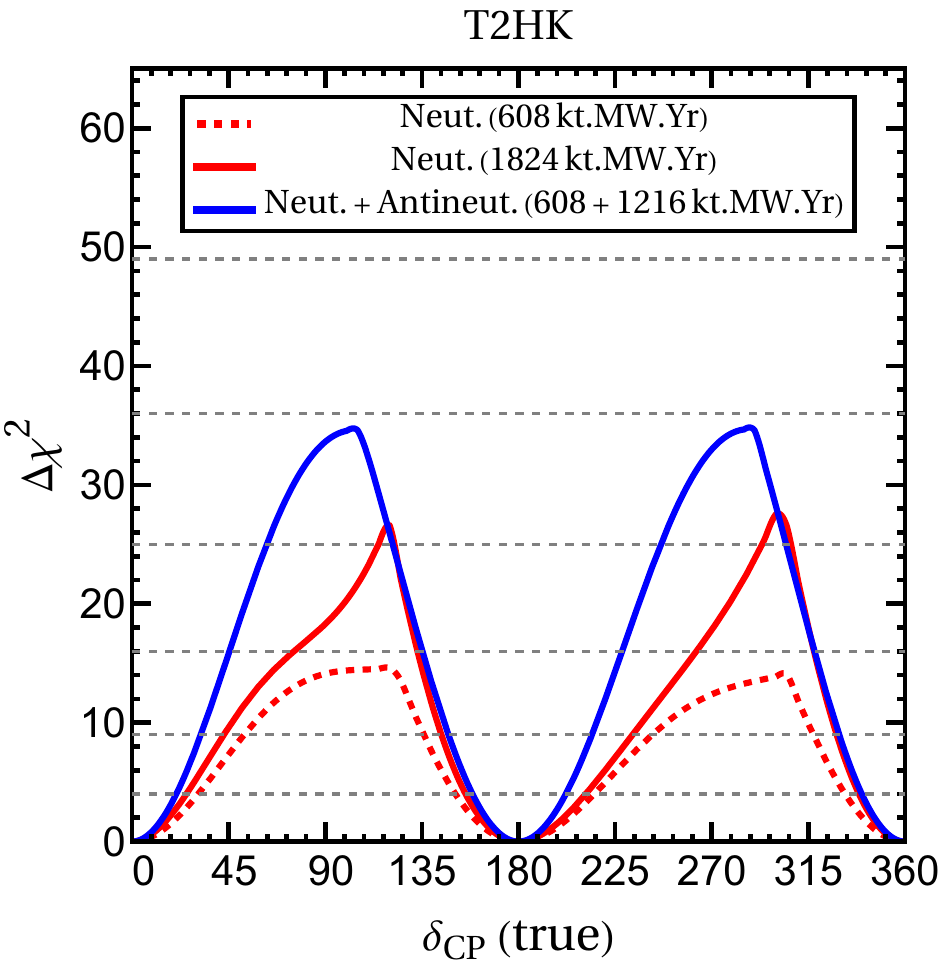}
    \mycaption{Sensitivity to $\delCP \neq 0,\pi$ from full spectral analysis for DUNE (left) and T2HK (right), for total exposures of 336~kt~MW~yr and 1824~kt~MW~yr, respectively. We compare the sensitivity obtained from using the full exposure in neutrino mode only (red) to the one when the exposure is divided into neutrino and anti-neutrino running with ratio 1:2 (blue). For T2HK we show for comparison also the neutrino-only sensitivity using 608~kt~MW~yr.}
    \label{fig:CP-full}
\end{figure}

\Cref{fig:CP-full} shows the DUNE and T2HK sensitivities to $\delCP$ based on the traditional method of fitting the full energy spectra. The figure compares the sensitivity when using the full exposure in neutrino mode only to the one when combining neutrino and anti-neutrino spectra dividing the exposure with ratio 1:2. For the full exposure we assume 336~kt~MW~yr for DUNE and 1824~kt~MW~yr for T2HK, respectively. This corresponds approximately to 7 years of data taking for both experiments.\footnote{Note that for DUNE we \emph{split} our default exposure of 336~kt~MW~yr from \cref{sec:formalism,sec:results} into neutrino and anti-neutrino running, whereas for T2HK we \emph{add} the anti-neutrino running to the 608~kt~MW~yr assumed above. For comparison we show for T2HK also sensitivity from our default neutrino-only exposure of 608~kt~MW~yr as dotted curve in \cref{fig:CP-full}.} We see that for DUNE the neutrino-only case provides better sensitivity, whereas for T2HK the neutrino/anti-neutrinos comparison is favourable. We interpret this result in the following way: thanks to the longer baseline as well as the broad spectrum covering first and second oscillation maxima in DUNE, in this case T violation (in the sense of an $L$-odd component in the oscillation probability) provides dominating information on $\delCP$. In the case of T2HK, due to the shorter baseline, the dominance of the first oscillation maximum and the relative smallness of matter effects, indeed the information from the CP-asymmetric observable is more powerful. 

\begin{figure}[t!]
    \centering
    \includegraphics[width=0.49\linewidth]{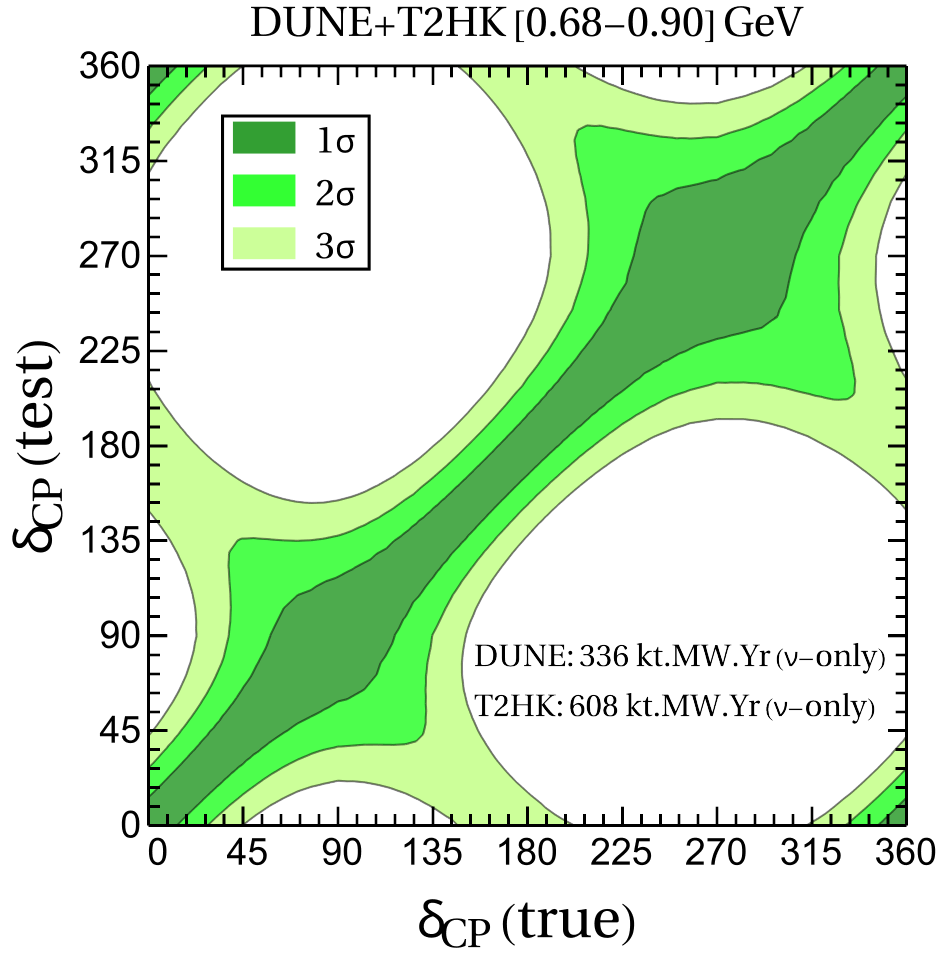}
     \includegraphics[width=0.49\linewidth]{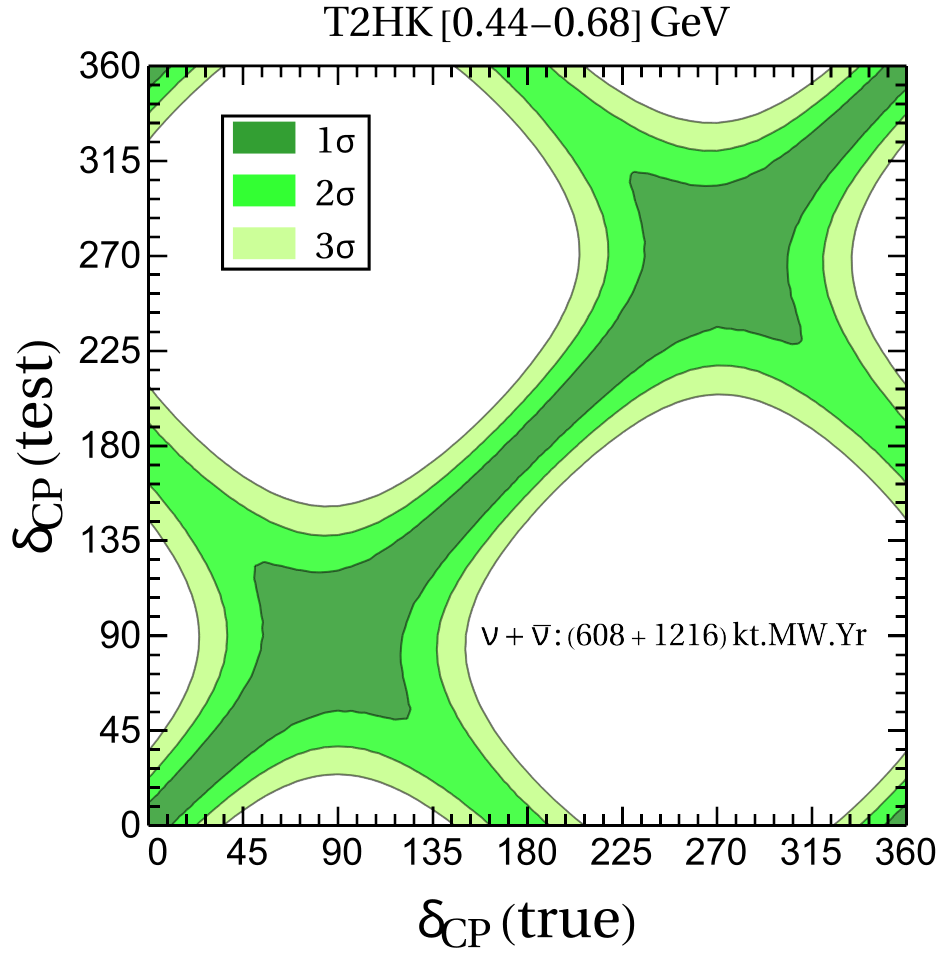}
    \mycaption{Accuracy of $\delCP^{\rm test}$ at $1,2,3\sigma$ versus $\delCP^{\rm true}$ for neutrino-only data DUNE+T2HK (combining two bins in the range $[0.68,0.92]$~GeV) and neutrino+anti-neutrino data from T2HK-only (two bins in $[0.44,0.68]$~GeV). Exposures are 336~kt~MW~yr for DUNE in neutrino mode only and 608/1216~kt~MW~yr for neutrino/anti-neutrino mode in T2HK.  These data can be accumulated in roughly 7 years runtime for both experiments.}
    \label{fig:TvsCP}
\end{figure}

In \cref{sec:bins} we show, that this result actually holds not only for the combined energy spectrum, but also for each individual energy bin for both experiments. We identify the two bins between 0.68 and 0.92~GeV as most sensitive to T violation from the DUNE/T2HK combination (see \cref{sec:results}) and the two bins between 0.44 and 0.68~GeV as most sensitive to CP violation in T2HK. In \cref{fig:TvsCP} we compare these two setups by studying the accuracy with which $\delCP$ can be determined. We observe from the figure that the two methods provide comparable sensitivity to $\delCP$, therefore allowing two independent determinations of $\delCP$ from T and CP violating observables, which of course, should lead to consistent results within statistical uncertainties. 

In this sense, the comparison of these independent determinations of $\delCP$ can be interpreted as a test of the CPT symmetry. We note however, that the ``CP analysis'' as performed here includes both, information from the $L$-dependence of the probability as well as from the neutrino + anti-neutrino comparison. In order to extract ``pure'' CP information, one would need to establish an observable depending only on the difference of the neutrino and anti-neutrino probabilities, which is, however, obscured due to environmental CP asymmetry introduced by the matter effect as well as by the relatively large neutrino component in the ``anti-neutrino'' beam mode. We leave such an investigation for future work.

\section{$X_T$ test}
\label{sec:XT}

The method to search for T violation used above is based on the presence of an $L$-odd component in the oscillation probability. As a variation of this, in our previous work \cite{Chatterjee:2024jzt} we have introduced a model-independent observable, called $X_T$ which is constructed from the difference of oscillation probabilities at two baselines. Here we briefly review this idea, specializing again to the standard three-flavour scenario and show numerical results.

We depart from the expression for the T-even probability, \cref{eq:Peven}. Under the assumption of T-conservation, the coefficients $c_i$ can be taken real and $P_{\rm odd}$ vanishes. With the abbreviations
\begin{equation}\label{eq:gamma}
  \begin{split}
  \gamma_i &= 4\sin^2\phi_{i1} \qquad (i=2,3) \,,\\
  \gamma_{23} &= 8 \sin\phi_{21}\sin\phi_{31}\cos(\phi_{31}-\phi_{21}) \,,
  \end{split}
\end{equation}  
the T-even part can be written as
\begin{align}\label{eq:Teven}
  P_{\rm even} =  \gamma_2 c_2^2  +
  \gamma_{23} c_2c_3 + \gamma_3 c_3^2
  \qquad (c_i\,\text{real})\,.
\end{align}
We consider now the difference of $P_{\rm even}$ at two baselines to define the quantity:
\begin{align} \label{eq:XTeven}
  X_T^{\rm even} \equiv P_{\rm even}(L_2) - P_{\rm even}(L_1) 
  =  \delta_2 c_2^2  + \delta_{23} c_2c_3 + \delta_3 c_3^2 
\end{align}
with 
\begin{align}\label{eq:delta}
  \delta_i \equiv \gamma_i(L_2) - \gamma_i(L_1) \quad (i=2,3,23) \,.
\end{align}

Without loss of generality we can assume $\delta_2>0$, as we have $\phi_{21} < \pi/2$ for T2HK and DUNE, which implies that $\delta_2 > 0$ for choosing $L_2 = L_{\rm DUNE} > L_1 = L_{\rm T2HK}$. The important observation is now that the right-hand side of \cref{eq:XTeven} is a \textit{non-negative function} of $c_2$ and $c_3$ if 
\begin{align}
  & \delta_3>0 \quad \text{and}\quad \delta_2>0 \,, \quad\text{and} \label{eq:cond1}\\
  & |\delta_{23}| < 2 \sqrt{\delta_2\delta_3} \,. \label{eq:cond2}
\end{align}
Hence, if these conditions are fulfilled and the observed value of the quantity $X_T$,
\begin{equation}\label{eq:XTobs}
  X_T^{\rm obs} = P_{\nu_\mu\to\nu_e}^{\rm obs}(L_2) - P_{\nu_\mu\to\nu_e}^{\rm obs}(L_1)
\end{equation}
is negative, T violation must be present in the fundamental theory~\cite{Chatterjee:2024jzt}.

\begin{figure}[t]
\centering
        {\includegraphics[width=0.48\textwidth]{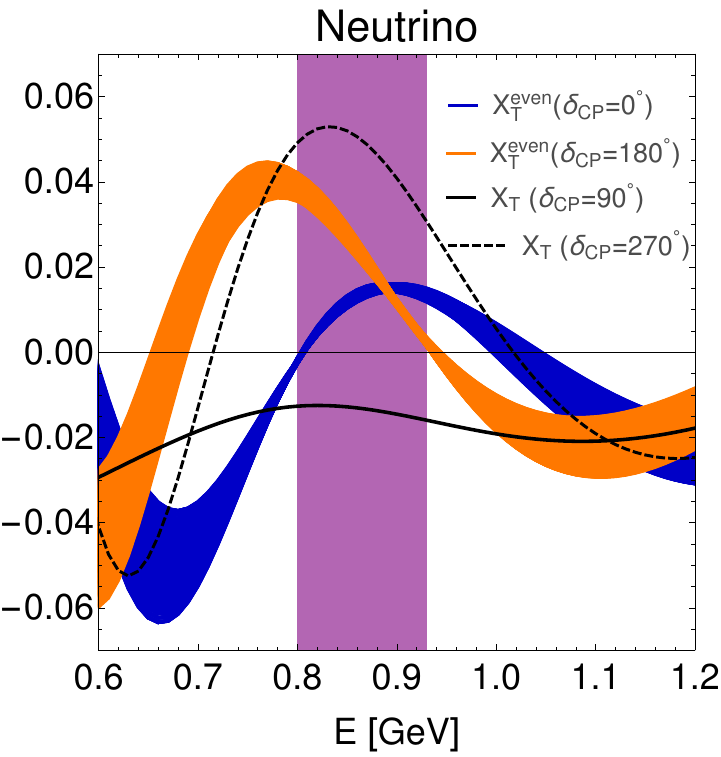}}
         {\includegraphics[width=0.48\textwidth]{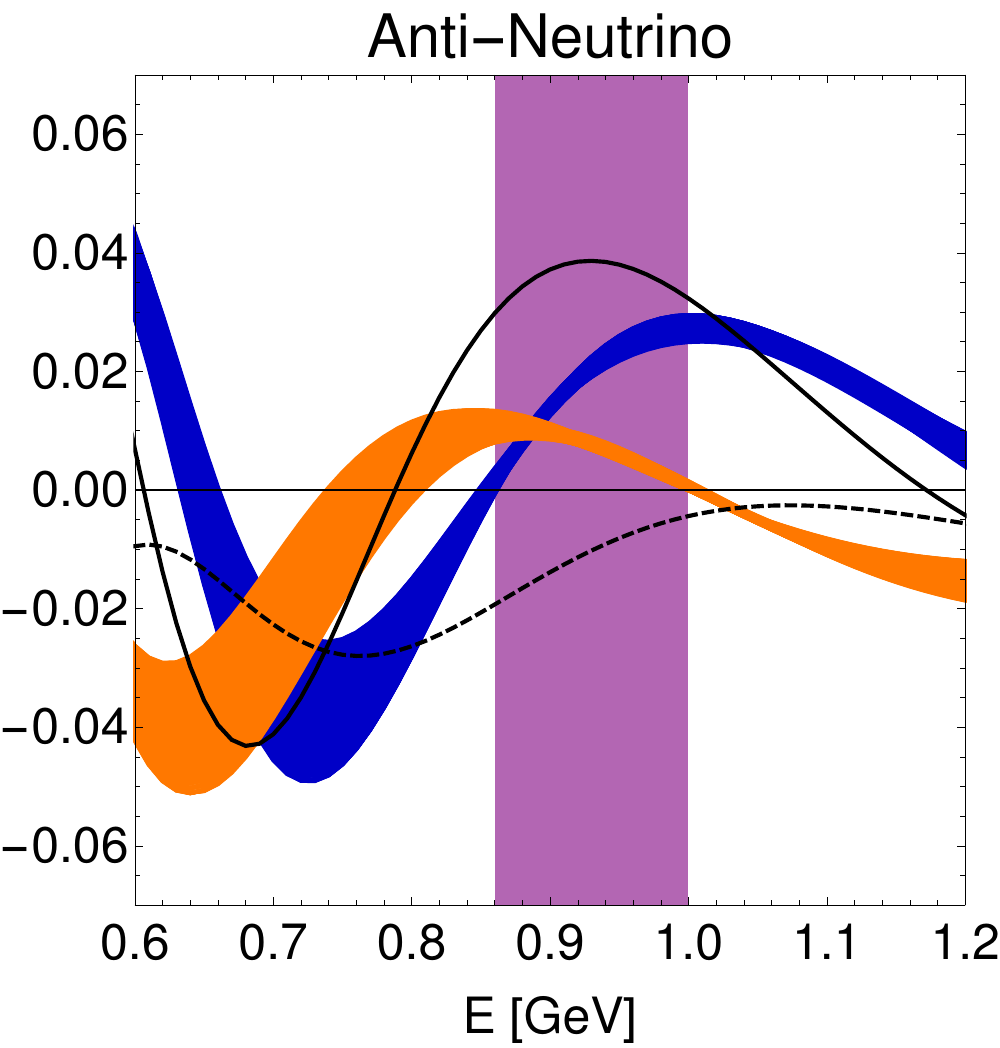}}
    \mycaption{$X_T = P_{\nu_\mu\to\nu_e}(L_{\rm DUNE})-P_{\nu_\mu\to\nu_e}(L_{\rm T2HK})$ as a function of neutrino energy for neutrinos (left) and anti-neutrinos (right). For the oscillation parameters the best values from \cite{Esteban:2024eli} have been used, assuming normal mass ordering. The orange and blue bands correspond to $X_T^{\rm even}$ implied by T-conservation, where the width of the band emerges from varying the lepton mixing angles within the current $3\sigma$ ranges. The solid-black (dashed-black) curves correspond to the case of maximal T violation with $\delta_{\rm CP} = 90^\circ (270^\circ)$. The purple-shaded band indicates the region where the conditions \cref{eq:cond1,eq:cond2} are fulfilled simultaneously.}
    \label{fig:XT-energy}
\end{figure}

In \cref{fig:XT-energy} we show $X_T^{\rm even}$ for the DUNE/T2HK combination as a function of neutrino energy, compared to the predicted observed value for $X_T$ if T-violation was maximal, i.e., for $\delta_{\rm CP} = 90^\circ (270^\circ)$. From the figure we observe that $X_T^{\rm even}$ is positive (both for $\delta_{\rm CP} = 0$ and $\pi$) in the energy interval where the conditions \cref{eq:cond1,eq:cond2} are fulfilled (indicated by the purple shading), as predicted by \cref{eq:XTeven}. The energy interval, where  \cref{eq:cond1,eq:cond2} are fulfilled is different for neutrinos and anti-neutrinos, because the effective mass-squared differences in matter are different (depending also on the assumed mass ordering): 
\begin{equation}\label{eq:bin} 
  \begin{split}
    E_\nu \in [0.80,0.92] \,{\rm GeV}\qquad \text{(neutrinos/NO and anti-neutrinos/IO)} \,, \\
    E_\nu \in [0.86,0.99] \,{\rm GeV}\qquad \text{(neutrinos/IO and anti-neutrinos/NO)} \,.
  \end{split}
\end{equation}
It is remarkable that the energy interval, where the conditions \cref{eq:cond1,eq:cond2} are fulfilled, agrees with one of the bins of the optimal sensitivity to the $L$-odd component discussed in \cref{sec:results}.

Furthermore, we observe from \cref{fig:XT-energy} that in the relevant energy range, $X_T$ for $\delta_{\rm CP} = 90^\circ(270^\circ)$ is negative (positive) for neutrinos, but positive (negative) for anti-neutrinos. Hence, using a neutrino beam, in the standard scenario this test can be applied for values of $\delta_{\rm CP}$ around $90^\circ$ \cite{Schwetz:2021cuj,Chatterjee:2024jzt}, where the T-violating value of $X_T$ has the opposite sign as for T conservation. Values around $270^\circ$ could be tested in principle with anti-neutrinos, although in practice the sensitivity is strongly reduced due to the large neutrino contamination in the ``anti-neutrino'' beam \cite{Chatterjee:2024jzt}. Therefore, we focus on the neutrino-beam mode for the numerical sensitivity analysis.

In our previous work \cite{Chatterjee:2024jzt} we developed a significance test based on the observable $X_T$. 
We applied this test without imposing any restriction on the coefficients $c_2$ and $c_3$, with the ambition of remaining as model-independent as possible. By marginalizing over 
$c_2$ and $c_3$, the right-hand side of \cref{eq:XTeven} can be made small and the significance of the test is given approximately by $|X_T^{\rm obs}|/\sigma_{X_T}$, with $\sigma_{X_T}$ denoting the uncertainty on $X_T^{\rm obs}$. In the present work we will take into account, that 
within the standard three-flavour framework the possible values of $c_2$ and $c_3$ are restricted by the allowed values of the oscillation parameters, implying that the right-hand side of \cref{eq:XTeven} will be constrained to sizeable non-zero values, increasing the significance to
\begin{equation}\label{eq:S_XT}
    S_{X_T} = \frac{\Theta(X_T^{\rm even} - X_T^{\rm obs})}{\sigma_{X_T}} \,.
\end{equation}
The $\Theta$-function takes into account that the $X_T$ test signals T-violation if $X_T^{\rm obs} < X_T^{\rm even}$, which follows from the form of \cref{eq:XTeven}. The uncertainty is estimated as follows:
\begin{equation}
    \label{eq:sigma_XT}
    \sigma_{X_T}^2 = \sum_x \frac{P_x^2}{S_x^2}\, 
    (S_x + B_x + \sigma_{xS}^2S_x^2 + \sigma_{xB}^2B_x^2)
    \,,
\end{equation}
where $x=$~DUNE,T2HK, $P_x$ is the corresponding oscillation probability, $S_x$ and $B_x$ are the number of signal and background events in experiment $x$, respectively, and $\sigma_{xS}\,(\sigma_{xB})$ is the relative systematic uncertainty in $S_x\, (B_x)$. 

In order to relate $X_T$ to observable quantities, we need to extract energy-averaged probabilities from observed event numbers. Hence, we define the averaged probability for a given energy bin as
\begin{equation}
    \langle P_{\nu_\mu\to\nu_e}^x\rangle \equiv P_x = \frac{S_x}{S_x(P_{\nu_\mu\to\nu_e}=1)}
\end{equation}
and use this in the calculation of $X_T$ as well as $\sigma_{X_T}$. Once the experiment has been performed, the probability needs to be extracted by considering $P^{\rm obs}_x = (N^{\rm obs}_x - B_x)/S_x(P_{\nu_\mu\to\nu_e}=1)$. Hence, it requires a prediction for the expected background events and the signal normalization, encoded in the expected number of signal events in the case of 100\% transition probability. The uncertainties on these are taken into account by $\sigma_{xB}$ and $\sigma_{xS}$ in \cref{eq:sigma_XT}.

The finite energy resolution will lead to a smearing of the probability between bins. This can lead to the situation that the conditions \cref{eq:cond1,eq:cond2} may no longer be fulfilled for energy-averaged quantities, when the chosen energy interval in~\cref{eq:bin} refers to the true neutrino energy. In order to avoid this effect, we assume for the following calculation an energy resolution for DUNE a factor of 2 better than our default configuration. (This corresponds to the green curve in \cref{fig:chi2-exposure}.) See also \cite{Chatterjee:2024jzt} for a discussion of energy resolution effects. Finally, for sensitivity estimates, there is an ambiguity in whether we use the assumed true event numbers or the fitted event numbers for $S_x$ and $P_x$ to calculate $\sigma_{X_T}$; both options are frequently used in the literature. Below we will show results for both approaches.

\begin{figure}[t]
\centering
\includegraphics[height=.48\textwidth]{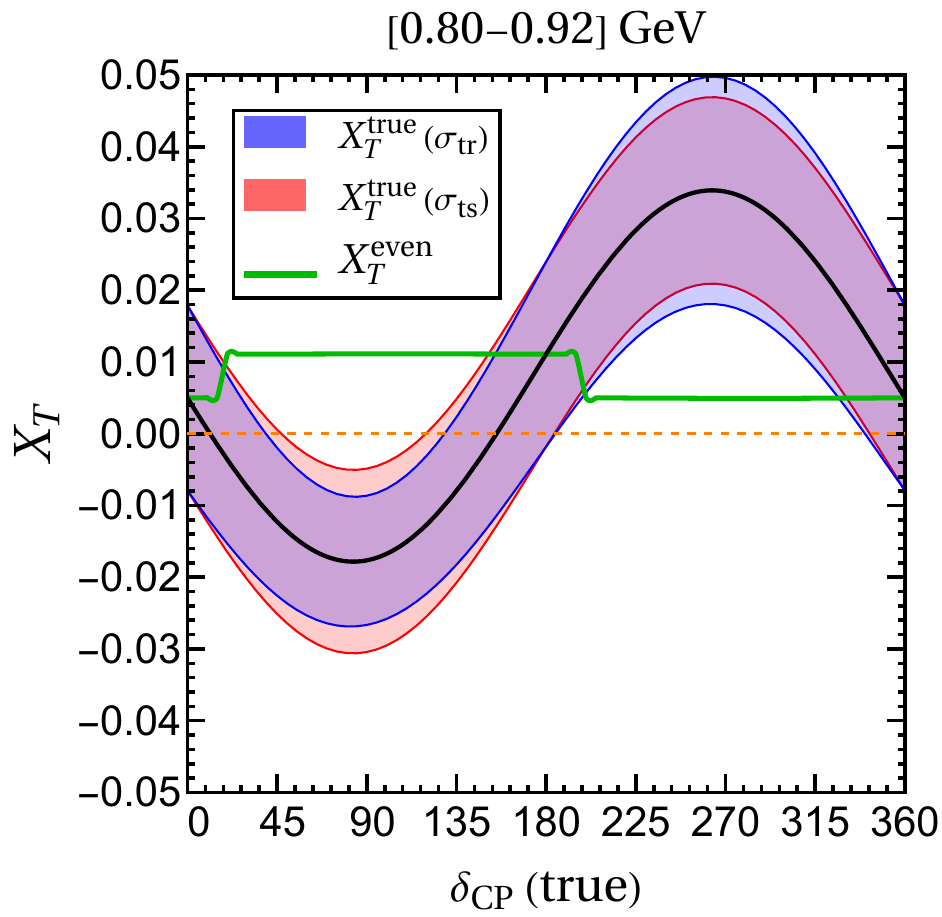}
\includegraphics[height=.48\textwidth]{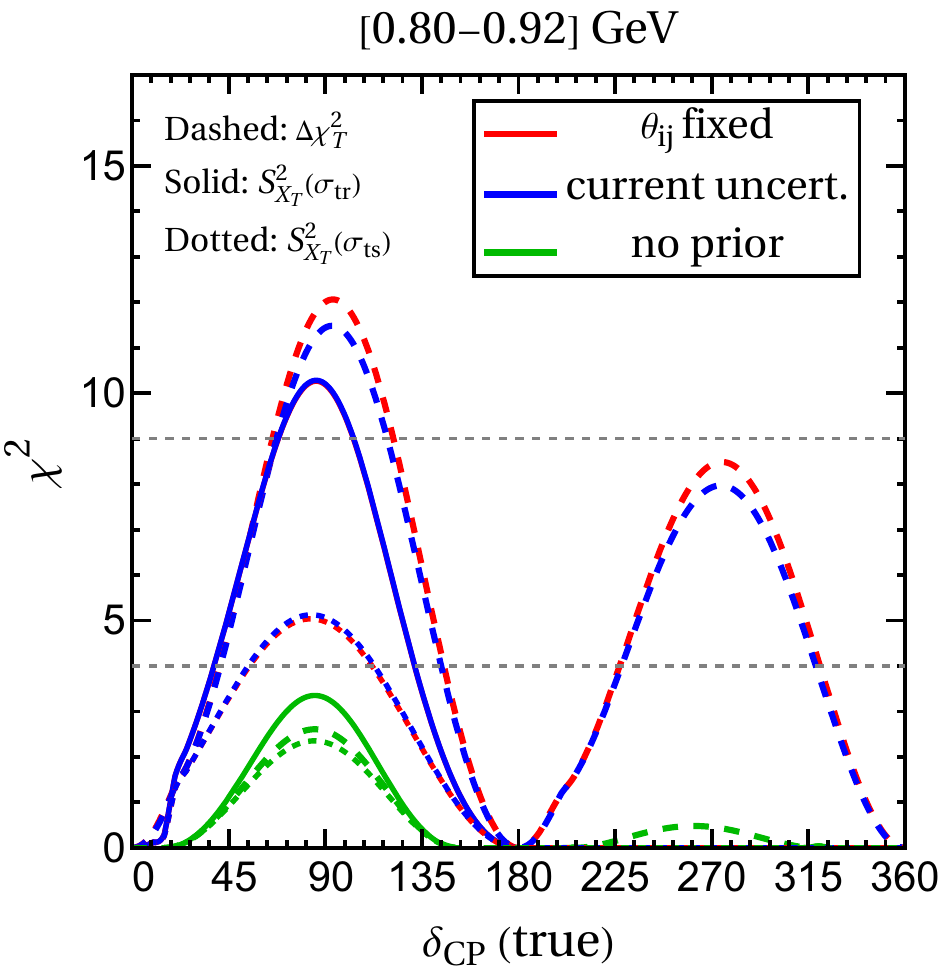}
\mycaption{Left: Predicted value of the observable $X_T \equiv  \langle P_{\nu_\mu\to\nu_e}^{\rm DUNE}\rangle  - \langle P_{\nu_\mu\to\nu_e}^{\rm T2HK}\rangle$ as a function of $\delCP$ (black curve). Error bands 
correspond to the $1\sigma$ uncertainty on $X_T$ according to our default exposures, where for the blue (red) band we use the true (best-fit) event rates, see \cref{eq:sigma_XT}. The green curve shows the predicted value of $X_T^{\rm even}$ at the best fit point for $\delCP = 0$ or $180^\circ$.  
Right: Sensitivity to $X_T$. The solid and dotted curves show the value of the significance $S^2_{X_T}$ as defined in \cref{eq:S_XT}, where for the solid (dotted) curves we use the true event rates (best-fit event rates) to calculate the uncertainty $\sigma_{X_T}$. For comparison, the dashed curves show $\Delta\chi^2$, \cref{eq:chi2T}. Colours indicate different assumptions about prior knowledge on mixing angles. The solid red and blue curves are practically overlapping. For both panels we assume an energy resolution in DUNE a factor 2 better than our default assumption.}
\label{fig:XT}
\end{figure}

In the left panel of \cref{fig:XT} we show the predicted values of $X_T$ as a function of the true value of $\delCP$ along with its uncertainty from \cref{eq:sigma_XT}, compared to the value of $X_T^{\rm even}$ (green curve) at the best-fit parameter values obtained from minimizing $\Delta\chi^2$ for the considered energy bin. In the right panel we show the corresponding significance based on \cref{eq:S_XT}. In the figure we compare the two options to calculate the uncertainty, of using true event numbers or best-fit event numbers. From the left panel we see that in the region where $X_T^{\rm true}< X_T^{\rm even}$ (i.e., where the $X_T$ test can be applied) uncertainties based on true event numbers are smaller and hence explaining the better sensitivity in that case visible in the right panel. In comparison with the $L$-dependence test discussed in \cref{sec:results}, we see a  reduced sensitivity of the $X_T$ test. This is expected, as it uses less information (only the difference of probabilities as single data point) as compared to at least two data points entering $\Delta\chi^2$.

The jump in the best-fit value of $X_T^{\rm even}$ in the left panel of \cref{fig:XT} emerges from switching between the two discrete cases $\delCP = 0,\pi$, which is visible also in a jump in the sensitivity in the right panel. As discussed above, the $X_T$ cannot be applied if $X_T^{\rm true} > X_T^{\rm even}$, which happens in the region $180^\circ < \delCP < 360^\circ$. Therefore,  the right panel displays sensitivity only for the $L$-odd test in that region of $\delCP$.

Let us emphasize the complementarity of a test based on $\Delta\chi^2$ as defined in \cref{eq:chi2T} and studied in \cref{sec:results} compared to the $X_T$ test. While the former does signal T-violation in the sense of the presence of an $L$-odd component, it does not automatically imply T-violating values of $X_T$. The analysis of \cref{sec:results} uses additional information (absolute probability values as well as bins where the conditions for the $X_T$ test are not satisfied), in contrast to the observable $X_T$, which contains reduced (though more model-independent) information corresponding to a \emph{single} data point.

\section{Conclusion}
\label{sec:conclusion}

In this work we have considered the sensitivity of the DUNE and T2HK long-baseline experiments to the complex phase in the PMNS mixing matrix, $\delCP$. In the standard three-flavour scenario considered here, fundamental T and CP violation are equivalent and both governed by $\delCP$. However, from a phenomenological perspective we distinguish here \emph{T violation} in the sense of an $L$-odd component of the transition probability (for a neutrino beam only) as contrasted to \emph{CP violation} from the comparison of neutrino and anti-neutrino transition probabilities. 

We show that the DUNE--T2HK combination offers excellent sensitivity to T violation by considering the event numbers in the energy range $E_\nu\in[0.68,0.92]$~GeV from neutrino beams only.
The sensitivity emerges due to high sensitivity from the second oscillation maximum at DUNE in the $[0.8,0.92]$~GeV energy bin, and a subtle interplay of DUNE and T2HK measurements in the $[0.68,0.8]$~GeV energy bin. For reasonable exposures, the sensitivity to T violation can reach more than $4\sigma$ in case of $\delCP \approx \pm 90^\circ$. We have studied the dependence of the sensitivity on the available exposure as well as on prior knowledge on the mixing angles, including the octant degeneracy of $\theta_{23}$, the latter playing an important role. 

When comparing T versus CP sensitivities, we find that DUNE has a better sensitivity to the former, whereas T2HK to the latter. We come to this conclusion by considering the sensitivity for a given exposure based on using the neutrino beam only as compared to splitting the same total exposure into neutrino and anti-neutrino beams in the ratio 1:2 (in order to obtain comparable event numbers). In that comparison we find for DUNE better sensitivity in the neutrino-only mode, whereas T2HK performs better in the split neutrino/anti-neutrino configuration. This statement holds when considering the full energy spectrum for both experiments as well as for each energy bin individually. We show that $\delCP$ can be independently determined from ``T-like'' and ``CP-like'' observables with comparable accuracy, by data available from DUNE and T2HK after about 7 years of data taking in neutrino mode only for DUNE and neutrino + anti-neutrino data from T2HK in the ratio 1:2. Such a comparison can be considered as a test of the CPT symmetry (note however the caveat mentioned at the end of \cref{sec:CP}).

Finally, we have considered the quantity $X_T$, which is given by the difference of transition probabilities in DUNE and T2HK. Under certain conditions, a negative value of $X_T$ is a model-independent signal of T violation. In \cref{sec:XT} we have studied the potential to establish T violation based on $X_T$ within the three-flavour framework. 

In conclusion, we offer an interpretation of future data from long-baseline experiments in terms of an $L$-odd component in the transition probability, which appears as a direct consequence of fundamental T violation. Such an analysis is based on neutrino data only and is complementary to the traditional method to search for CP violation in comparing neutrino versus anti-neutrino-enhanced event spectra. Within the standard three-flavour framework, we have found that the DUNE+T2HK combination offers excellent sensitivity to T violation based on realistic exposures. The low-energy part of the DUNE event spectrum is crucial for this analysis. 

\subsection*{Acknowledgements}

The work of S.S.C.\ is funded by the Deutsche Forschungsgemeinschaft (DFG, German Research Foundation), project number 510963981. K.S.\ acknowledges the financial support received from the bi-nationally supervised doctoral degrees program of the German Academic Exchange Service (DAAD). S.P.\ would like to acknowledge the funding support from SERB, Government of India, under MATRICS
project with grant no.\ MTR/2023/000687.

\appendix

\section{Energy dependence of T and CP sensitivities}
\label{sec:bins}

\begin{figure}[t!]
\centering
       \includegraphics[width=5.4cm]{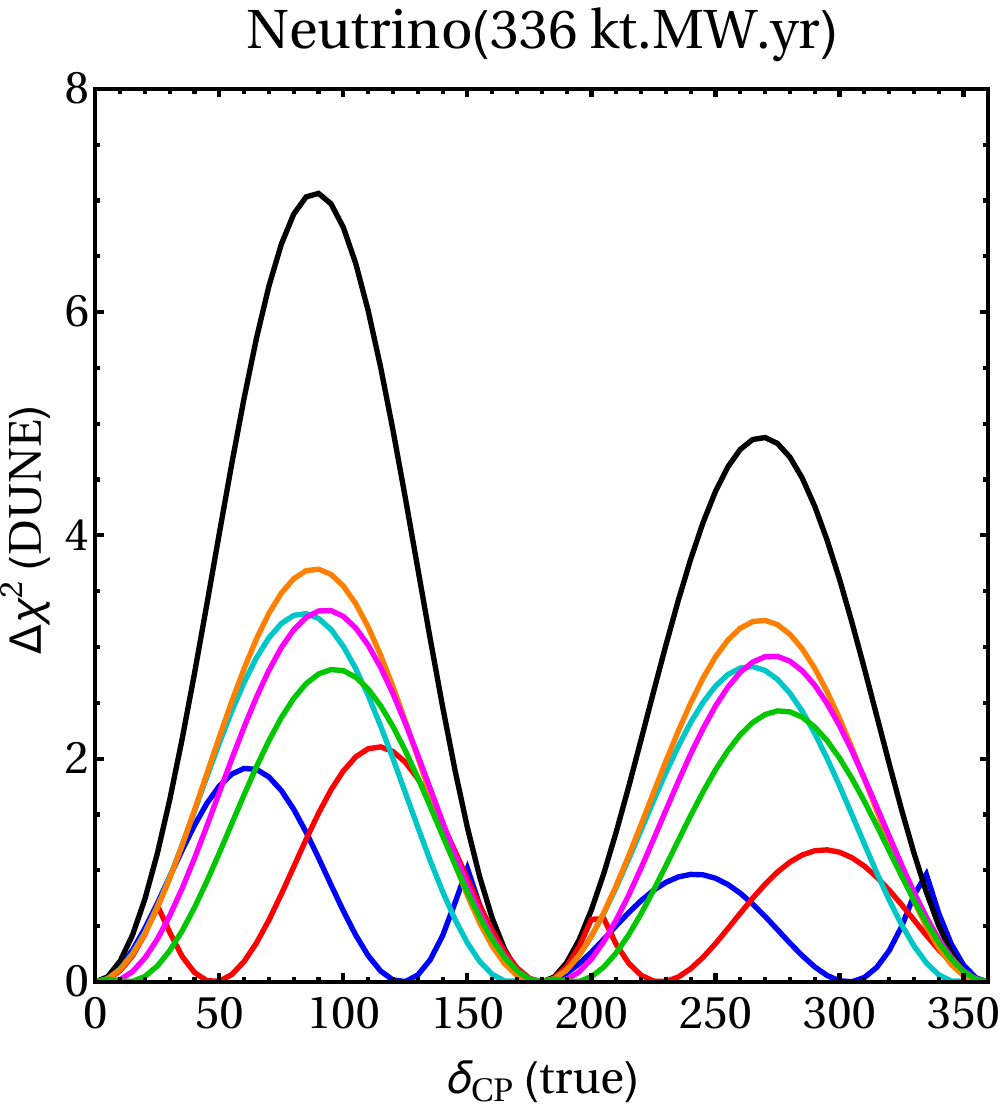}
       \qquad
        \includegraphics[width=5.4cm]{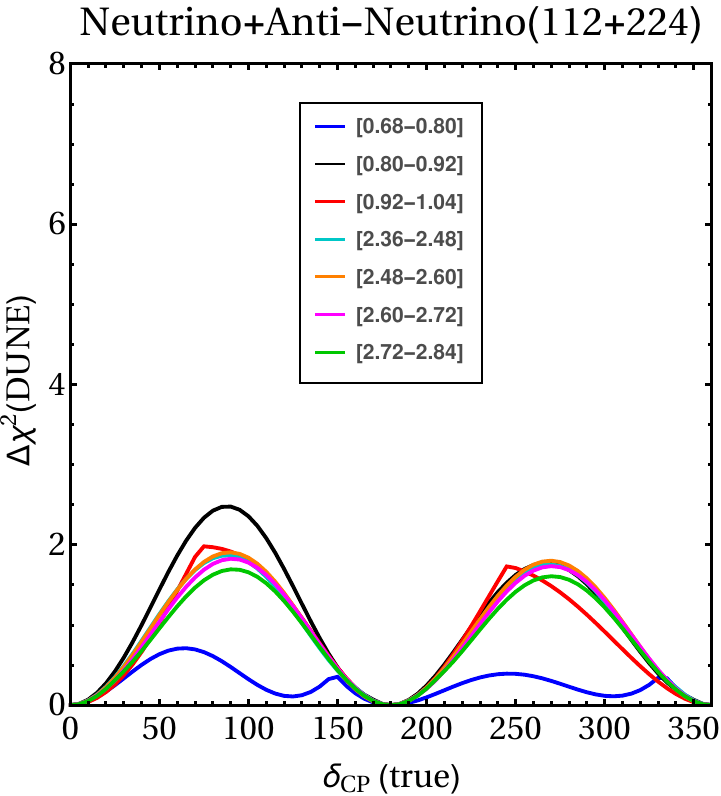}
    \mycaption{Sensitivity to T-violating values of $\delCP$ for individual energy bins in DUNE. The left  panel corresponds to neutrino-beam mode only for 336~kt~MW~yr. For the right panel we split the same exposure into neutrino and anti-neutrino beam running with ratio 1:2.}
    \label{fig:bins-DUNE}
\end{figure}

\begin{figure}[t!]
       \includegraphics[height=5.45cm,width=5.4cm]{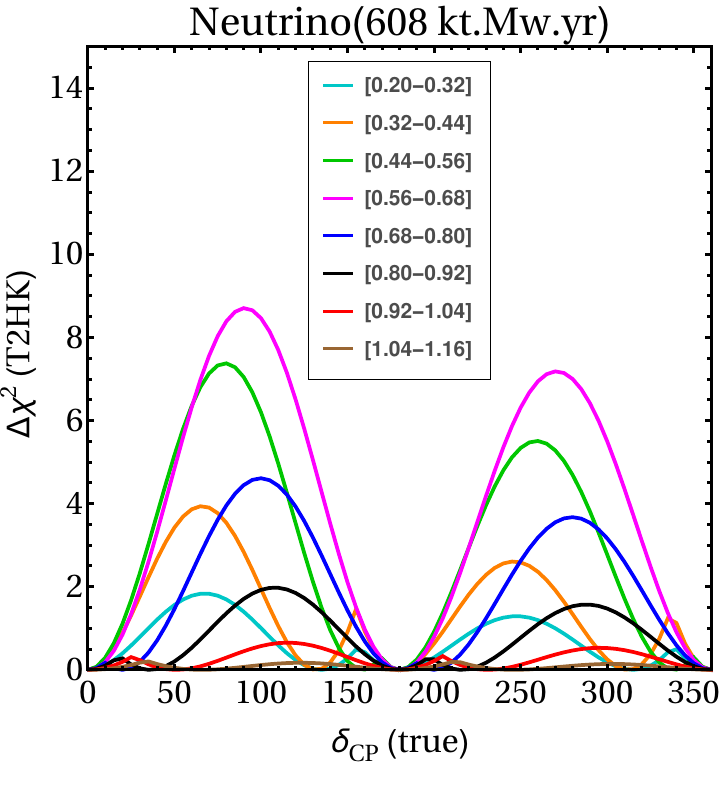}
         \includegraphics[height=5.45cm,width=5.4cm]{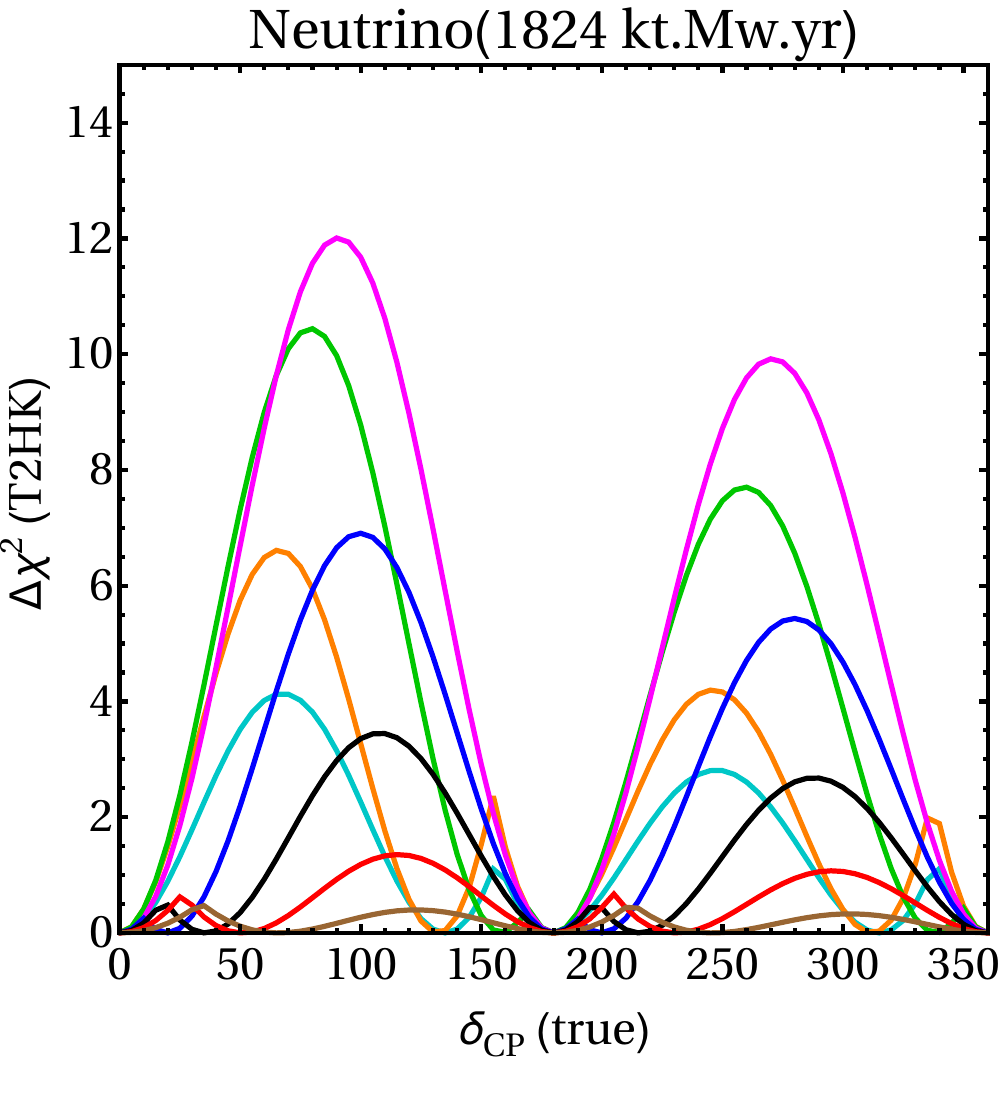}
           \includegraphics[height=5.45cm,width=5.4cm]{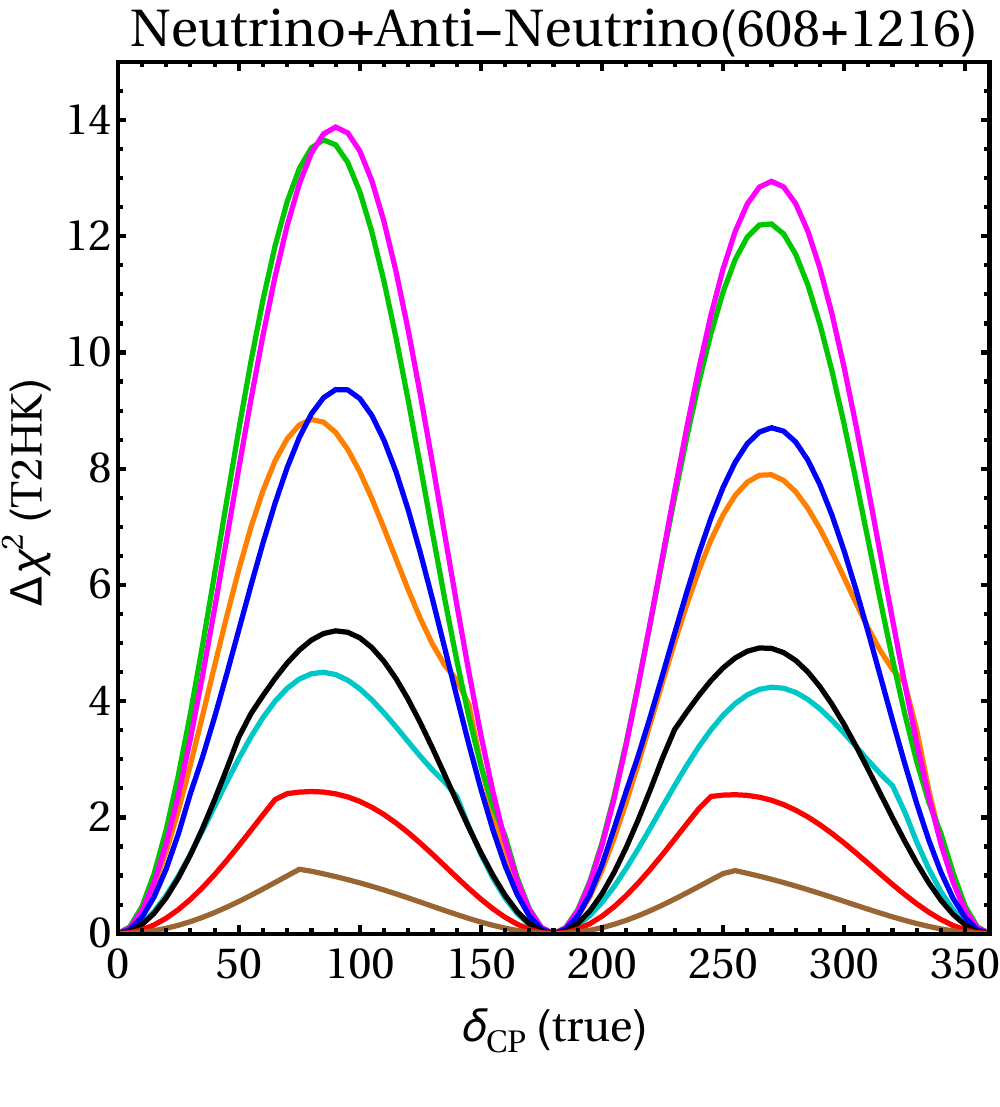}
    \mycaption{Sensitivity to T-violating values of $\delCP$ for individual energy bins in T2HK. Left and middel panels correspond to neutrino-beam mode only for 608 and 1824~kt~MW~yr, respectively. For the right panel we combine neutrino and anti-neutrino beam running with ratio 1:2 for 1824~kt~MW~yr in total.}
    \label{fig:bins-T2HK}
\end{figure}

In this appendix we study in some detail the bin-wise sensitivity of the DUNE and T2HK experiments and compare the sensitivity of neutrino-only data (``T violation'') to the one from the neutrino/anti-neutrino combination (``CP violation''). These results are summarised in \cref{fig:bins-DUNE,fig:bins-T2HK} for the DUNE and the T2HK experiments, respectively.

The left panel of \cref{fig:bins-DUNE} shows the DUNE sensitivity to T violation (neutrino data only) for a wide range of energy bins. We see that for the chosen exposure, several bins provide sensitivity close to $2\sigma$, but the bin of 
$[0.80,0.92]$~GeV (shown in black) clearly sticks out with a sensitivity well above $2\sigma$. To gain further understanding of this result, we can consider an approximate expression for the transition probability in constant matter~\cite{Cervera:2000kp,Freund:2001pn,Akhmedov:2004ny}:
\begin{align}
P_{\nu_\mu\to\nu_e} &\approx \sin^22\theta_{13} \,\sin^2\theta_{23} \, \frac{\sin^2
         \Delta (1 - A)}{(1 - A)^2} +
         \alpha^2 \sin^2 2\theta_{12} \, \cos^2\theta_{23} \cos^2\theta_{13}
         \frac{\sin^2 A\Delta}{A^2} \nonumber\\
         &+ \alpha \, \cos\theta_{13}\sin 2\theta_{13} \, \sin 2\theta_{12} \,
         \sin2\theta_{23} \cos(\Delta + \delta_{\rm CP}) \, 
	 \frac{\sin\Delta A}{A} \, \frac{\sin \Delta (1 - A)}{1-A} \,,
\end{align}
with the definitions
\begin{equation}
\Delta \equiv \frac{\Delta m^2_{31} L}{4E_\nu} \,,\quad
A \equiv  \frac{2E_\nu V}{\Delta m^2_{31}} \,, \quad
\alpha \equiv \frac{\Delta m^2_{21}}{\Delta m^2_{31}} \,,
\end{equation}
where $V$ is the
effective matter potential~\cite{Wolfenstein:1977ue}. Hence, the amplitude of the T-odd contribution, i.e., the coefficient in front of $\sin\delCP$, is given by
\begin{equation}\label{eq:Aodd}
         \mathcal{A}_{\rm odd}  = -\alpha \, \cos\theta_{13}\sin 2\theta_{13} \, \sin 2\theta_{12} \,
         \sin2\theta_{23} \sin\Delta \, 
	 \frac{\sin\Delta A}{A} \, \frac{\sin \Delta (1 - A)}{1-A} \,.
\end{equation}

\begin{figure}[t!]
\centering
\includegraphics[width=0.5\textwidth]{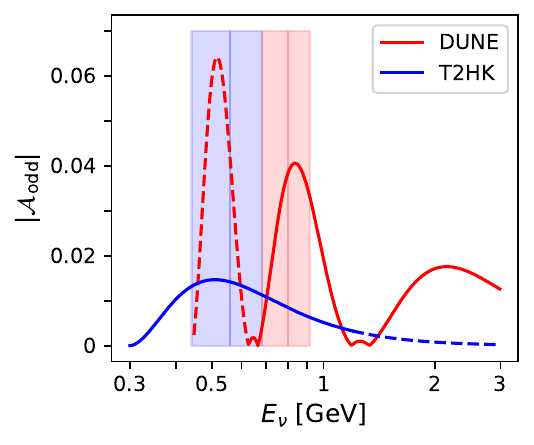}
\mycaption{Amplitude of the T-odd part of the transition probability $P_{\nu_\mu\to\nu_e}$, see \cref{eq:Aodd}, as a function of neutrino energy at the DUNE and T2HK baselines. The dashed parts of the curves indicate very small event numbers. Oscillation parameters are set to the values in \cref{tab:osc-params}. The energy bins with the optimal sensitivity to T (CP) violation are marked by the red (blue) shaded region.}
\label{fig:Aodd}
\end{figure}

The modulus of this quantity is shown in \cref{fig:Aodd} for the DUNE and T2HK baselines. We clearly see that the T-odd amplitude has a sharp maximum for DUNE precisely around the two optimal energy bins identified above, hence, explaining the high sensitivity to T violation of these bins, see also the bi-probability plots in \cref{fig:bi-prob}.\footnote{DUNE has practically no signal events at the very low energies of the third maximum around 0.5~GeV, which however, is close to the optimal energy for T2HK.} 

We remark that when three bins around the 1st oscillation maximum, from 2.36 to 2.72~GeV are combined, a similar sensitivity is obtained as from the bin at the 2nd oscillation maximum, see \cref{fig:bins-DUNE}. Note however, in the energy range 
$[2.36, 2.72]$~GeV, we obtain between 280 ($\delCP\approx 90^\circ$) and 429 ($\delCP\approx 270^\circ$) events in DUNE for our assumed exposure, whereas in the 
$[0.80,0.92]$~GeV bin only between 7 ($\delCP\approx 90^\circ$) and 32 ($\delCP \approx 270^\circ$) events are expected. This shows a much higher ``sensitivity per event'' at the 2nd than at the 1st oscillation maximum.

Comparing the left and right panels of \cref{fig:bins-DUNE}, we observe that in DUNE most bins give a comparable contribution to $\Delta\chi^2$ in the CP-violation analysis, in contrast to the single ``high-sensitivity'' bin for the T-violation analysis. Note that for each individual bin, the sensitivity for neutrino-only data is higher than when splitting the same exposure into neutrino and anti-neutrino running, supporting our interpretation of better T than CP sensitivity of DUNE.

Moving now to T2HK shown in \cref{fig:bins-T2HK}, we find also in this case very good sensitivity based on neutrino data only (left and middle panels). However, the comparison of the middle and right panels of \cref{fig:bins-T2HK} shows the opposite behaviour as for DUNE:  for each single bin in T2HK the sensitivity to CP violation using neutrino and anti-neutrino is higher than using neutrino-data only for the same total exposure. Furthermore, we observe a clear trend of the sensitivity being dominated by bins close the first maximum of the CP-odd amplitude, and we identify the two bins $[0.44, 0.56]$ and $[0.56,0.68]$~GeV as the most sensitive bins when considering T2HK on its own, compare \cref{fig:Aodd}. 

These results justify the choice of the energy intervals used for comparing the sensitivity to $\delCP$ from T-odd observables 
($[0.68,0.92]$~GeV) to the ones to CP-odd observables 
($[0.44,0.68]$~GeV) in \cref{sec:CP}.

\section{Estimating corrections for variable matter density}
\label{sec:density}

All results in this paper assume a constant, line-averaged matter density of $\bar{\rho} = 2.84 \,\rm g/cm^3$ for both T2HK and DUNE. While this matches the DUNE baseline value, it differs slightly from the T2HK average of $2.6 \,\rm g/cm^3$. In reality, both experiments feature non-constant, asymmetric density profiles~\cite{Hagiwara:2011kw, Roe:2017zdw}. This appendix details the impact of matter density on CP and T violation sensitivities by comparing two scenarios:
(I) constant but different averaged densities for each experiment, and
(II) the real, variable density profiles along each baseline.

\begin{figure}[t!]
\centering
\includegraphics[height=.48\textwidth]{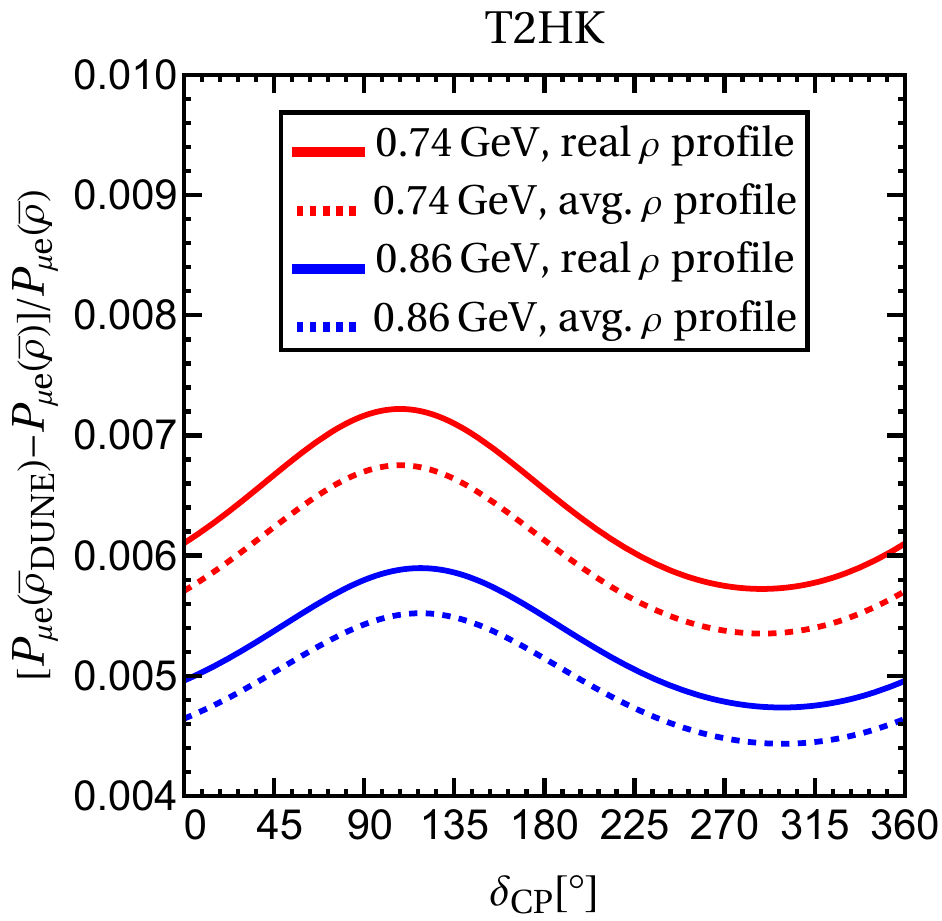}
\includegraphics[height=.48\textwidth]{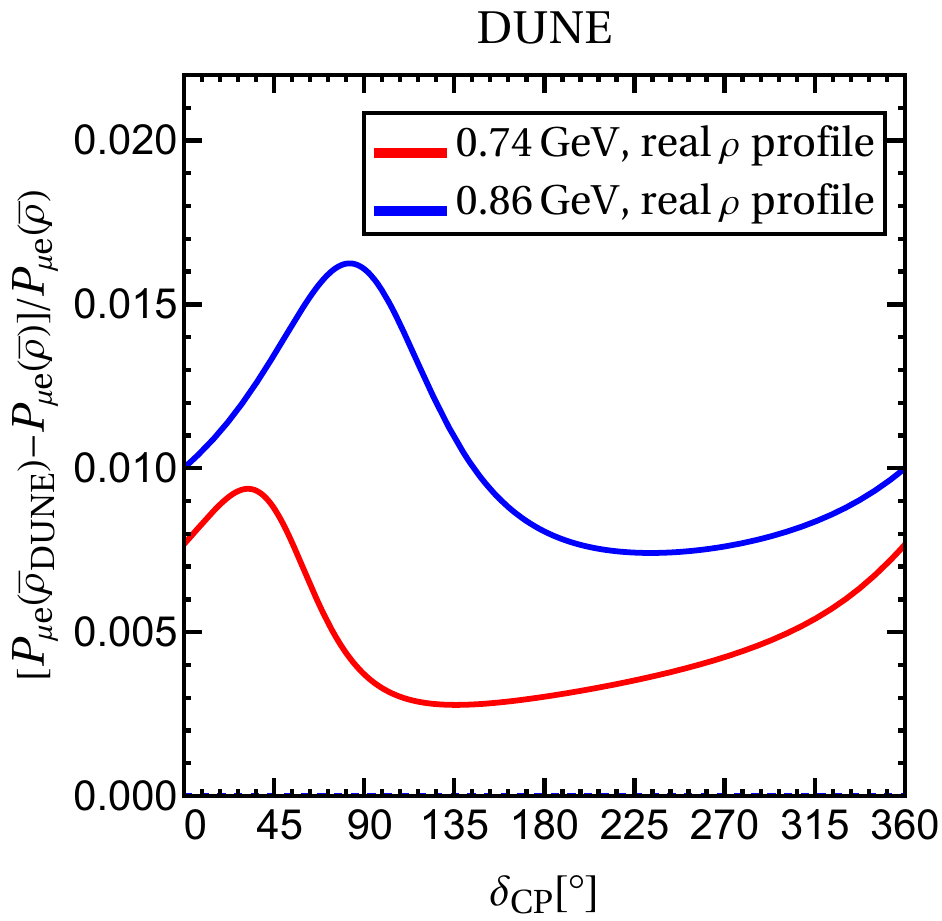}
\mycaption{Relative corrections to the $\nu_\mu \rightarrow \nu_e$ appearance probability ($P_{\mu e}$) for T2HK (left) and DUNE (right) as a function of the CP-phase $\delta_{\rm CP}$. Results are shown for two energies, $E_\nu = 0.74$ and $0.86$~GeV. Solid lines represent the realistic non-constant density profiles~\cite{Hagiwara:2011kw, Roe:2017zdw}, while dashed lines show the effect of using different constant average densities.}
\label{fig:density}
\end{figure}

\Cref{fig:density} shows the relative error in $P_{\mu e}$ induced by these density choices. For T2HK (left panel, dashed lines), using the DUNE-averaged density ($\bar{\rho}_{\rm DUNE} = 2.84 \,\rm g/cm^3$) instead of its own ($\bar{\rho}_{\rm T2HK} = 2.6 \,\rm g/cm^3$) results in a negligible error of less than $1\%$ across all $\delta_{\rm CP}$ values. This corresponds to a change of less than one event per hundred in a typical bin, justifying its neglect in our sensitivity analysis. This case is omitted for DUNE (right panel) as the relative error is zero by definition when $\bar{\rho}_{\rm DUNE} = \bar{\rho}$.

We now consider the realistic profiles from~\cite{Hagiwara:2011kw, Roe:2017zdw} (Case II, solid lines). Despite T2HK's step-like asymmetric potential and DUNE's close to symmetric profile, the resulting corrections remain remarkably small. The error consistently stays below $1\%$, with a single exception: for DUNE at $E_{\nu} = 0.86$ GeV and $\delta_{\rm CP} \approx 90^\circ$, it reaches $\sim1.5\%$. These deviations are negligible in terms of expected event counts.

In conclusion, the use of a unified constant density of $2.84 \,\rm g/cm^3$ is well-justified for the comparative analysis of T2HK and DUNE presented in this work, as the introduced errors are completely insignificant compared to statistical uncertainties, see e.g., \cref{fig:bi-prob} for the typical size of error bars in the relevant bins.

\bibliographystyle{JHEP_improved}
\bibliography{refs}

\end{document}